\documentclass{arxiv}
\usepackage{amsmath}
\usepackage{subfigure}

\usepackage{etoolbox}
\usepackage{microtype}
\usepackage{ragged2e}
\usepackage{comment}

\usepackage[square,numbers]{natbib} 

\patchcmd{\abstract}{\noindent}{\noindent\justifying}{}{}

\AtBeginDocument{%
  \justifying
  \emergencystretch=2em
}

\usepackage[justification=justified,singlelinecheck=false]{caption}
\usepackage[normalem]{ulem}

\begin{document}

\title{Strongly interacting matter with criticality induced by modified excluded volume in core-collapse supernova simulations}

\author{Anil~Kumar$^{1}$\orcid{0000-0001-5833-7595}, 
Noshad~Khosravi~Largani$^{1,2,3}$\orcid{0000-0003-1551-0508},
Stefan~Typel$^{4,5,6}$\orcid{0000-0003-3238-9973},
Pablo~Cerd{\'a}-Dur{\'a}n$^{7,8}$\orcid{0000-0003-4293-340X},
\\
Alejandro~Torres-Forn{\'e}$^{7,8}$\orcid{0000-0001-8709-5118},
and
Tobias~Fischer$^{1,3}$\orcid{0000-0003-2479-344X}}

\affil{$^1$~Institute of Theoretical Physics, Wrocław University of Science and Technology, Wybrzeże Wyspiańskiego 27, 50-370 Wrocław, Poland}

\affil{$^2$~Institute of Theoretical Physics, University of Wroclaw, Pl. M. Borna 9, 50-204 Wroclaw, Poland}

\affil{$^3$~Research Center for Computational Physics and Data Processing, Institute of Physics, Silesian University in Opava, Bezručovo nám. 13, CZ-746-01 Opava, Czech Republic}

\affil{$^4$~Institut für Kernphysik, Technische Universität Darmstadt, Fachbereich Physik, Schlossgartenstraße 9, 64289 Darmstadt, Germany}

\affil{$^5$~GSI Helmholtzzentrum für Schwerionenforschung, Theorie, Planckstraße 1, 64291 Darmstadt, Germany}

\affil{$^6$~Helmholtz Forschungsakademie Hessen für FAIR (HFHF), GSI Helmholtzzentrum für Schwerionenforschung, Campus Darmstadt,
Schlossgartenstraße 2, 64289 Darmstadt, Germany}

\affil{$^7$~Departament d’Astronomia i Astrof{\'i}sica, Universitat de Val{\'e}ncia, C/ Dr Moliner 50, 46100, Burjassot (Val{\'e}ncia), Spain}

\affil{$^8$~Observatori Astronòmic, Universitat de València, E-46980, Paterna (València), Spain}

\email{anil.1@iitj.ac.in, s.typel@gsi.de, bert-tobias.fischer@pwr.edu.pl}


\begin{abstract}
\noindent\justifying
This article reviews critically the core-collapse supernova explosion mechanism associated with a sufficiently strong first-order phase transition from normal nuclear, in general hadronic matter to deconfined quark matter, which commonly assumes Gibbs` conditions for the coexistence of phases and a phase transition construction accordingly. 
To this end, a novel class of multi-purpose equation of state (EOS) is developed, based on the modified excluded volume (MEV) approach employing a medium-dependent excluded-volume functional within the relativistic mean field framework with density-dependent meson-nucleon couplings.
The chosen MEV parametrisation features the change in the number of degrees of freedom, mimicking the EOS softening in excess of nuclear saturation density, featuring a first-order phase transition with van der Waals like behaviour and the presence of a critical point at high temperatures.
Simulations of core-collapse supernovae are performed, based on general relativistic neutrino radiation hydrodynamics in spherical symmetry, in order to explore the previously reported supernova explosion scenario within this class of phenomenological modified microscopic hadronic EOS. A burst-like neutrino signature is released, substantially longer than previously reported based on common hadron-quark hybrid model EOS with two-phase approach and Gibbs' phase-transition construction, as observable signal, which is complemented by a gravitational wave mode analysis.
\end{abstract}

\section{Introduction}
A core-collapse supernova (CCSN) is one of the most energetic transients in the current universe, marking the end of a massive star's life. 
Electron captures, first on ion group nuclei and later also on unbound protons trigger the substantial pressure loss that results in an adiabatic collapse of the iron core eventually, reaching supersonic velocities. 
When densities are reached in the collapsing stellar core, on the order of nuclear saturation density ($n_0$), the collapse halts due to the repulsive nuclear force, and the core bounces back with the formation of a shock wave.
The latter propagates outwards rapidly, however, suffering from the loss of energy due to the photodisintegration of still infalling heavy nuclei.
A second loss occurs when the shock wave passes across the electron neutrinospheres of last inelastic scattering, which results in the release of the $\nu_e$ deleptonization burst and causes the expanding bounce shock to stall, turning into an accretion front.
The revival of the stalled bounce shock, primarily due to the transfer of energy from the central protoneutron star (PNS)---it forms at core bounce and contains about 99\% of the gravitational binding energy gained from the stellar core collapse---is the subject of the CCSN explosion mechanism.
Several scenarios have been studied theoretically, the magneto-rotational mechanism in the presence of highly magnetized and rotating stellar cores~\cite{Bisnovatyi-Kogan70,Bisnovatyi-Kogan93} and via neutrino heating and cooling~\cite{Bethe85}.
A third mechanism has been brought foreword immediately after the SN1987A, due to a phase transition from nuclear matter to deconfined quark matter~\cite{TakaharaSato1988PThPh80}. The idea has been revived recently, employing microscopic quark matter model equations of state (EOS)~\cite{Fischer18,Khosravi:2024ApJ964}.
The latter CCSN explosion mechanism is the subject of investigation of this article
(for reviews about core-collapse supernova phenomenology, c.f. Refs.~\cite{Mezzacappa05,Janka07,Janka12,Janka2025ARNPS75}, and references therein).

It is evident that one of the largest uncertainties in modeling CCSN is the high-density EOS \cite{Fischer17}.
Especially first-order phase transitions have long been studied in the context of CCSN, namely, the gas-liquid phase transition below and around nuclear saturation density and a possible phase transition to deconfined quark matter at suprasaturation densities. 
The first transition can be understood from first principles of nuclear physics, in particular the dependence on temperature with the presence of a critical point at around $T\simeq 10$--20~MeV~(c.f. Ref.~\cite{Fischer2020PhRvC102} and references therein). 
The latter, on the other hand, is entirely model-dependent due to the unknown physical mechanism of (de)confinement. 
While ab initio lattice QCD provides a robust prediction of a cross over transition at a pseudo-critical temperature of about $156\pm 1.5$~MeV, at vanishing baryon density~\cite{Bazavov:2014,Borsanyi:2014,Bazavov:2019}, at finite and large baryon density phenomenological quark matter models have long been employed for astrophysical studies, such as the thermodynamic bag models \cite{Farhi84,Sagert09}, with the extension including repulsive vector interactions \cite{Klaehn:2015,Klaehn:2017}, Nambu--Jona-Lasino models~\cite{NJL:1961,Klevansky1992RvMP64,Buballa2005PhR407} and quark matter models employing the relativistic density functional (RDF) approach~\cite{Kaltenborn17,Bastian:2021} based on the classical string-flip model~\cite{Goddard1973NuPhB56_stringflip,JohnsonThorn1976PhRvD13_stringflip,Horowitz1985PhRvD31_stringflip,Ropke86}. It has been realized recently that, despite the latter models implement a phenomenological approach to confinement, however, the resulting density dependence cannot be related to QCD~\cite{ShuklaPok2025arXiv250706741S,ShuklaPok2025JSPC300058S}, despite reaching the causal limit at finite density, depending on the string-flip parametrization.
Furthermore, all such hadron-quark hybrid EOS suffer from the two-phase approach, i.e. different hadronic/nuclear matter and quark matter models, which results in a first-order phase transition by design, usually without a critical point at a finite density and temperature. 
This has important consequences for the CCSN explosion mechanism, as it depends on details of the phase transition construction, e.g., the onset density of quark matter, the density jump before reaching the pure quark matter phase and the pressure slope in between the two stable phases, which make it difficult to relate to quark matter model parameters within microscopic calculations~\cite{Kaltenborn17,Bastian:2021}.

Special emphasis has been devoted to EOS constraints from the high-precision pulsar mass measurements of about $2~M_\odot$~\cite{Fonseca:2021}. It relates seemingly to a rather stiff behavior of the EOS at supersaturation density \cite{Klaehn2006PhRvC74}, which is complemented by the analysis of gravitational waves (GW) detected from the very first confirmed binary neutron star merger event GW170817, pointing towards not too large neutron star radii on the order of $9$--$13$~km for neutron stars of about $1.25$--$1.5~M_\odot$ \cite{Abbott17PhRvL,Lattimer18PhRvL}.
This is partly in tension with results from the NICER NASA mission~\cite{NICER_Miller2019,NICER_Watts2019,NICER_Miller2021,NICER_Riley2021}.

In this article we aim to overcome the caveats of the two-phase approach, by adopting the modified excluded-volume (MEV) approach of Ref.~\cite{Typel2018Univ4}, which is based on the DD2 relativistic mean field (RMF) approach for infinite nuclear matter with density dependent meson-nucleon couplings.
This MEV model modulates the canonical EOS behavior at supersaturation densities, in such a way that it enables a continuous 
phase transition within the DD2 RMF framework, mimicking the change of the number of degrees of freedom. 
The selection of parameters employed in the present study features the behavior of van der Waals gases, with over- and under-critical regimes as well as the presence of a critical endpoint of the first-order phase transition at high temperatures. 
Furthermore, we compare the resulting CCSN evolution with a classical phase transition construction assuming Gibbs conditions of phase equilibrium. 
The results contain details that will enable us to distinguish one of the cases versus the other in both neutrinos and potentially in GWs.
For the latter, a detailed GW mode analysis is performed, based on the leading-order general relativistic perturbative approach of astroseismology~\cite{Thorne81,Torres-Forne2019MNRAS482}. 
Particularly interesting is the finding that for the case with phase transition construction successful CCSN explosions can no longer be obtained, due to the too small density jump, e.g., compared to the class of RDF hybrid EOS of Ref.~\cite{Bastian:2021}. This puts severe constraints on the hadron-quark hybrid EOS parameters for this CCSN explosion mechanism to operate. 

The manuscript is organized as follows.
In Sec.~\ref{sec:eos} we will discuss the novel class od DD2-MEV EOS, for which we briefly revisit the DD2 RMF model and the implementation of the MEV approach. 
CCSN simulation results with the DD2-MEV EOS will be discussed in Sec.~\ref{sec:sim}, with the results being analyzed with respect to potential detection prospects, the neutrino signal in Sec.~\ref{sec:neutrino} and a gravitational wave mode analysis in Sec.~\ref{sec:modes}.
The manuscript closes with a summary in Sec.~\ref{sec:summary}.

\section{Equation of State}
\label{sec:eos}
In the following we will revisit the DD2 RMF model with density-dependent couplings and the further extension to the MEV approach, which enables us to study the passage from the low-density to the high-density region with a first-order phase-transition construction as well as a continuous change through an instability region.
Special emphasis will be given to the rearrangement terms, that arise due to the medium dependence of the imposed MEV functional.

\subsection{DD2 density-dependent RMF EOS revisited}
Within the DD2 RMF model with density-dependent nucleon-meson couplings, nucleons (neutrons and protons) interact through the exchange of meson fields--scalar meson ($\sigma$), vector meson ($\omega$) and iso-vector vector meson ($\rho$). 
The latter is responsible for explicit isospin-asymmetry. 

The Lagrangian density for this model is,
\begin{align}\label{eqn:lagrangian}
\mathcal{L} & = \sum_{N} 
\bar{\psi}_{N} (i\gamma_{\mu} D_{N}^{\mu} - {M}_{N}^{\ast})\psi_{N} 
+ \frac{1}{2}(\partial_{\mu}\sigma\partial^{\mu}\sigma - m_{\sigma}^2{\sigma}^2)\nonumber\\ 
&  -  \frac{1}{4}\omega_{\mu\nu}\omega^{\mu\nu} + \frac{1}{2}m_{\omega}^2\omega_{\mu}\omega^{\mu} {- \frac{1}{4}\boldsymbol{\rho}_{\mu\nu}\cdot\boldsymbol{\rho}^{\mu\nu}}
+\frac{1}{2}{m_{\rho}^2}\boldsymbol{{\rho}_{\mu}}\cdot\boldsymbol{{\rho}^{\mu}},
\end{align}
with the covariant derivative, $iD_{N}^{\mu} = i\partial^\mu - \Gamma_{\omega} \omega^\mu - \Gamma_{\rho} \boldsymbol{\tau} \cdot \boldsymbol{\rho}^{\mu}$, and the effective mass operator, 
$M_{N}^{\ast}=m_{N}-\Gamma_{\sigma} \sigma$, including the minimal coupling terms. 
The sum over $N$ includes neutrons ($n$) and protons ($p$) as formal 
hadronic degrees of freedom.
The coupling parameters $\Gamma_{\omega}$, $\Gamma_{\rho}$, and
$\Gamma_{\sigma}$ are assumed to be density-dependent to capture the density evolution of nuclear interaction, as introduced in Ref.~\cite{TYPEL1999331}. 
The parameters of these functions in the DD2 model are given in Ref.~\cite{Typel:2009sy}.
The nucleon wave functions, $\psi_{N,\,k}$, of momentum $k$, are solutions of the Dirac equation,
\begin{align}
    \left[\boldsymbol{{\alpha} \cdot k} + \beta(m_{N} - S_{N}) + V_{N} \right] \psi_{N,\, k}(\vec{r}) = E_{N k} \psi_{N,\, k}(\vec{r})
\end{align}
where $S_{N}$ and $V_{N}$ are the scalar and vector potentials,
\begin{align}
   S_{N} &= \Gamma_{\sigma}\sigma + S^r \nonumber\\
   V_{N} &= \Gamma_{\omega} N\omega_0 + \Gamma_{\rho} \tau_{3N} \rho_0 + V^r
\end{align}
with the scalar and vector rearrangement terms given as,
\begin{equation}\label{eqn:rearrange}
\begin{aligned}
S^r &=  \sum_{N} \left[
\frac{\partial \Gamma_{\sigma}}{\partial n^{(s)}} \sigma n_{N}^{(s)} - \frac{\partial \Gamma_{\rho}}{\partial n^{(s)}} \rho \boldsymbol{\tau} n_{N}^{(v)} - \frac{\partial \Gamma_{\omega}}{\partial n^{(s)}} \omega_{0}n^{(v)} \right]~, \\
V^r &= \sum_{N} \left[ \frac{\partial \Gamma_{\omega}}{\partial n^{(v)}}\omega_{0}n_{N}^{(v)} + \frac{\partial \Gamma_{\rho}}{\partial n_{N}^{(v)}} \rho \boldsymbol{\tau} n_{N}^{(v)} - \frac{\partial \Gamma_{\sigma}}{\partial n^{(v)}} \sigma n_{N}^{(s)} \right]~.
\end{aligned}
\end{equation}
These rearrangement terms arise as correction in the energy functional due to the density-dependent couplings. They are required for the thermodynamic consistency of the theory.
However, in the DD2 model, the coupling parameters are independent of the scalar density. Hence, the scalar rearrangement term $S^{r}$ does not contribute. 
The rearrangement terms are functions of the total scalar density, $n^{(s)}={\sum}_N n^{(s)}_N$, with
\begin{align}
n^{(s)}_N = g_{N} \int \frac{d^3k}{(2\pi)^3}\frac{m^*_N}{\sqrt{k^2+{m^*_N}^2}}\left[f(E^*_N;\{T,\mu^*_N\}) + f(E^*_N;\{T,-\mu^*_N\}) \right]~,
\label{eq:ns}
\end{align}
with spin degeneracy, $g_N=2J_N+1$
and total vector or baryonic density, $n^{(v)}=n_B={\sum}_N n^{(v)}_N$,
\begin{align}
n^{(v)}_N = g_{N} \int \frac{d^3k}{(2\pi)^3}
\left[f(E^*_N;\{T,\mu^*_N\}) - f(E^*_N,\{T,-\mu^*_N\})\right]~.
\label{eq:nv}
\end{align}
Here, $f(E^*_N,\{T,\mu^*_N\})$ is the nucleon Fermi-Dirac distribution function,
\begin{align}
    f(E^*_N,\{T,\mu^*_N\}) = \frac{1}{1+\exp\{(E^*_N-\mu^*_N)/T\}}~,
\end{align}
with the gap equations for the effective mass of nucleons,
\begin{equation}
m_{N}^{*}=m_{N}-S_{N}
\label{eq:mass-gap}
\end{equation}
appearing explicitly in the dispersion relations, 
$E_N^*(k)=\sqrt{k^2 + m^*_N}$, and the effective chemical potentials,
\begin{equation}
\mu^*_N = \mu_N - V_N~.
\label{eq:mu-gap}
\end{equation}
The chemical potentials of neutrons and protons can be expressed as
\begin{equation}
\mu_{n} = \mu_{\rm B} \qquad \mbox{and} \qquad \mu_{p} = \mu_{\rm B} + \mu_{Q}
\end{equation}
with the baryon and charge chemical potentials, $\mu_{\rm B}$ and $\mu_{Q}$, respectively.
The pressure can be obtained from the energy-momentum tensor, which includes contributions from meson fields and the vector rearrangement term as follows,
\begin{align}
\label{eqn:pressure}
P &= \sum_N p_N + p_{\mathrm{meson}} + p^r = 
\sum_{N} p_{N}
- \frac{1}{2}{m_{\sigma}^2\sigma^2} + \frac{1}{2}m_{\omega}^2\omega_0^2 + \frac{1}{2}m_{\rho}^2\rho_{0}^2 + n^{(v)}V^r
\end{align}
with the partial pressures
\begin{align}
\label{eqn:partial_pressure}
 p_{N} = 
\frac{g_N}{3}\int \frac{d^3k}{(2\pi)^3}\frac{k^2}{E^*_N}
\left[f(E^*_N;\{T,\mu^*_N\}) + f(E^*_N,\{T,-\mu^*_N\})\right]~.
\end{align}
The remaining quantities are then determined through the usual thermodynamic relations, in terms of pressure derivatives.
For practical purposes, we invert the problem numerically, in order to arrive at the pressure in terms of temperature $T$, baryon density $n_{\rm B}$ and isospin asymmetry, the latter expressed in terms of $Y_p:=n_p/n_{\rm B}$, we employ a numerical root finding procedure in these independent quantities to solve the gap equations~\eqref{eq:mass-gap} and \eqref{eq:mu-gap}. 

\begin{table*}[t!]
\centering
\caption{Excluded volume parametrisation as well as bulk and neutron star properties.}
\begin{tabular}{ccccccc}
\hline
\hline
$v\,\,\,[{\rm fm}]$ & $\mathcal{S}$ & $T_0\,\,\,[{\rm MeV}]$ & $n_{\rm cut}\,\,\,[{\rm fm}^{-3}]$ & $\rho_{\rm onset}^a\,\,\,[{\rm fm}^{-3}]$ & $\rho_{\rm end}^b\,\,\,[{\rm fm}^{-3}]$ & $M_{\rm max}^c\,\,\,[{\rm M}_\odot]$ \\
\hline
2&3 &270 & 0.149$^d$ &0.268 & 0.412& 2.19\\
\hline
\end{tabular}
\\
\vspace{2mm} 
\raggedright $^a$~onset density for the phase transition at $T=1$~MeV and $Y_p=0.3$ \\
\raggedright $^b$~end density for phase transition at $T=1$~MeV and $Y_p=0.3$ \\
\raggedright $^c$~maximum neutron star mass ($T=0$ and $\beta$-equilibrium) \\
\raggedright $^d$~this value corresponds to the DD2 saturation density
\label{tab:eos}
\end{table*}

\subsection{Phase transition via excluded volume approach}
The excluded-volume mechanism has been incorporated into the RMF formalism in order to describe the dissolution of nuclei at high densities \citep[c.f. Refs.][]{Typel10,HS}. It is assumed that these clusters of nucleons have a finite size which reduces their available volume for the thermal motion leading to an effective repulsion and an increase of the pressure. 
This approach can be generalized by choosing modified functional forms of the available volume fractions and applying the formalism to other degrees of freedom, e.g., nucleons. In the MEV approach, the interpretation changes from the geometric picture to that of a change of the effective number of degrees of freedom. 
Thus the effects are modeled with effective degeneracy factors or effective potentials \cite{2016EPJA...52...16T} that allow to describe a stiffening or softening of EoS  \cite{2015A&A...577A..40B}.

The available volume fraction of nucleons suggested for nuclear matter in Gaussian form in reference \cite{2016EPJA...52...16T} does not depend on temperature. 
Here we consider the modified available-volume fraction function for protons and neutrons
\begin{equation}
\Phi_N(T,x) = 1 + \mathcal{S}\,g_1(T)\,\Theta(x)\,\exp\left(-\frac{1}{2x^2}\right)~,
\end{equation}
depending on temperature $T$ and on the quantity $x$, defined as,
\begin{align}
x = v\,(n_p^{(v)}+n_n^{(v)}-g_8(T)\,n_{cut})~,
\end{align}
where $g_1$ and $g_8$ are functions of temperature,
\begin{align}\label{eqn:temp_functions}
g_t(T) = \Theta(T_{0}-T)\,\exp\left[-\frac{t}{2}\left(\frac{T}{T_{0}-T}\right)^2\right]~,
\end{align}
as suggested in Ref.~\cite{Typel2018Univ4}. Here, $\Theta$ is the heavyside step function, $\mathcal{S}$ and $v$ are free parameters for the excluded volume.
The remaining parameters, $n_{cut}$ and $T_0$, are determined by boundary conditions. 
The values of all parameters are given in Table~\ref{tab:eos}. 

In order to implement the MEV mechanism in the RMF approach, the degeneracy factors, $g_{N}$, which appear in the calculation of the scalar and vector densities, expressions~\eqref{eq:ns} and \eqref{eq:nv}, are replaced by effective couplings as follows,
\begin{align}
g_N\longrightarrow g_{\text{eff}}(T,n_{\rm B}) = g_N\,\Phi_N(T,n_{\rm B})~,
\end{align}
depending explicitly on the baryon density in the current parametrization, since $n_{\rm B}=n_p^{(v)}+n_n^{(v)}$.
An additional contribution appears in Eq.~\eqref{eqn:rearrange} for the rearrangement terms, as follows,
\begin{equation}
S^r \longrightarrow S^r + S^r_\Phi~,
\quad {\rm with}\quad 
S^r_\Phi = -\sum_N p_N \frac{\partial \ln \Phi_N}{\partial n^{(s)}_N}~,
\end{equation}
and further,
\begin{equation}
V^r \longrightarrow V^r + V^r_\Phi~, 
\quad {\rm with}\quad 
V^r_\Phi = \sum_Np_N\frac{\partial\ln{\Phi_N}}{{\partial}n^{(v)}_N}~,
\end{equation}
with partial nucleon pressures, $p_N$, as given in Eq.~\eqref{eqn:partial_pressure}.

\begin{figure}[t!]
\centering
\subfigure[~$T=1$~MeV and varying values of $Y_p$]{
\includegraphics[width=0.49\textwidth]{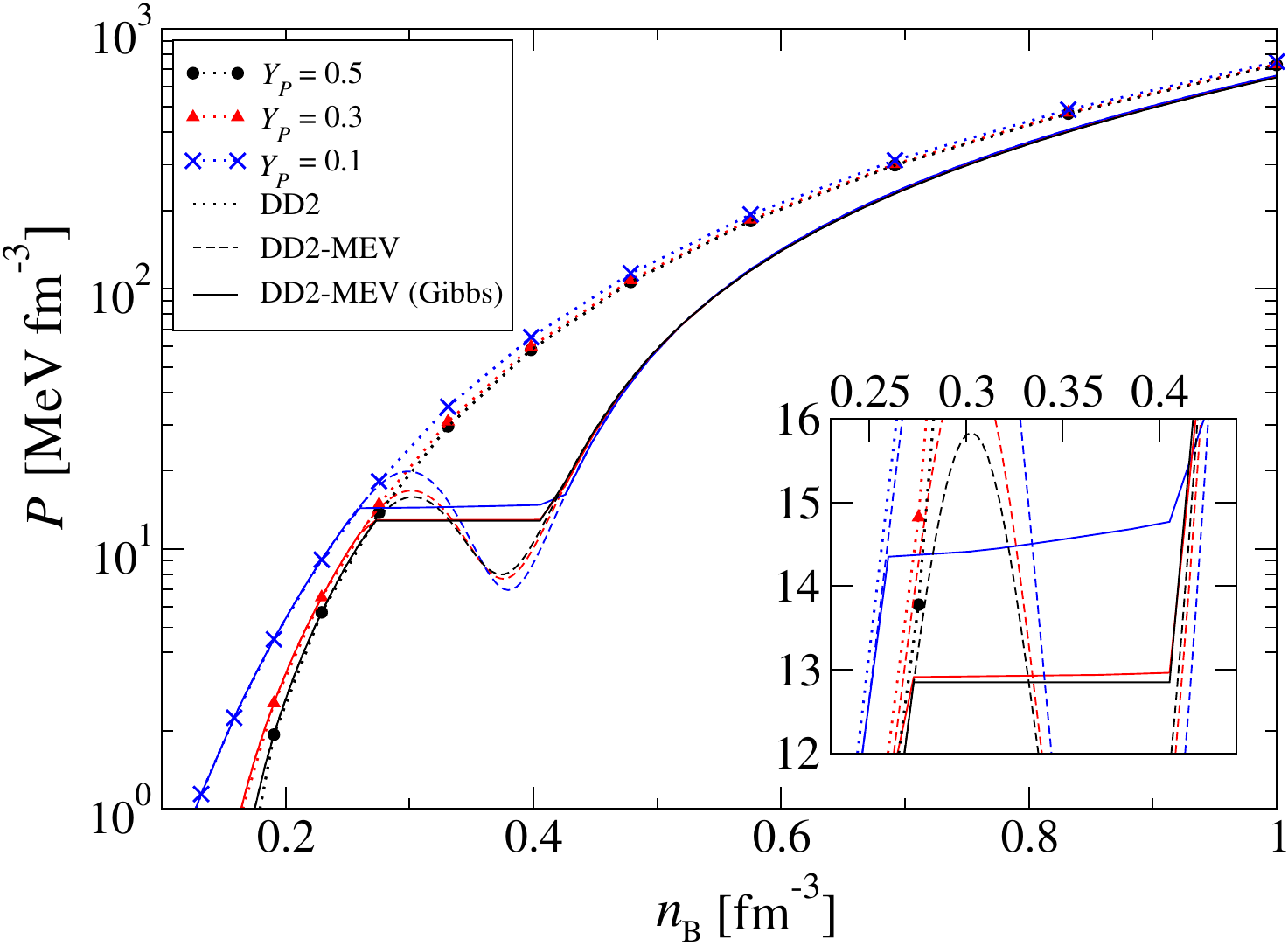}
\label{fig:EOS_a}
}
\hfill
\hspace{-0.75cm}
\subfigure[~$Y_p=0.3$ and three different temperatures]{
\includegraphics[width=0.495\textwidth]{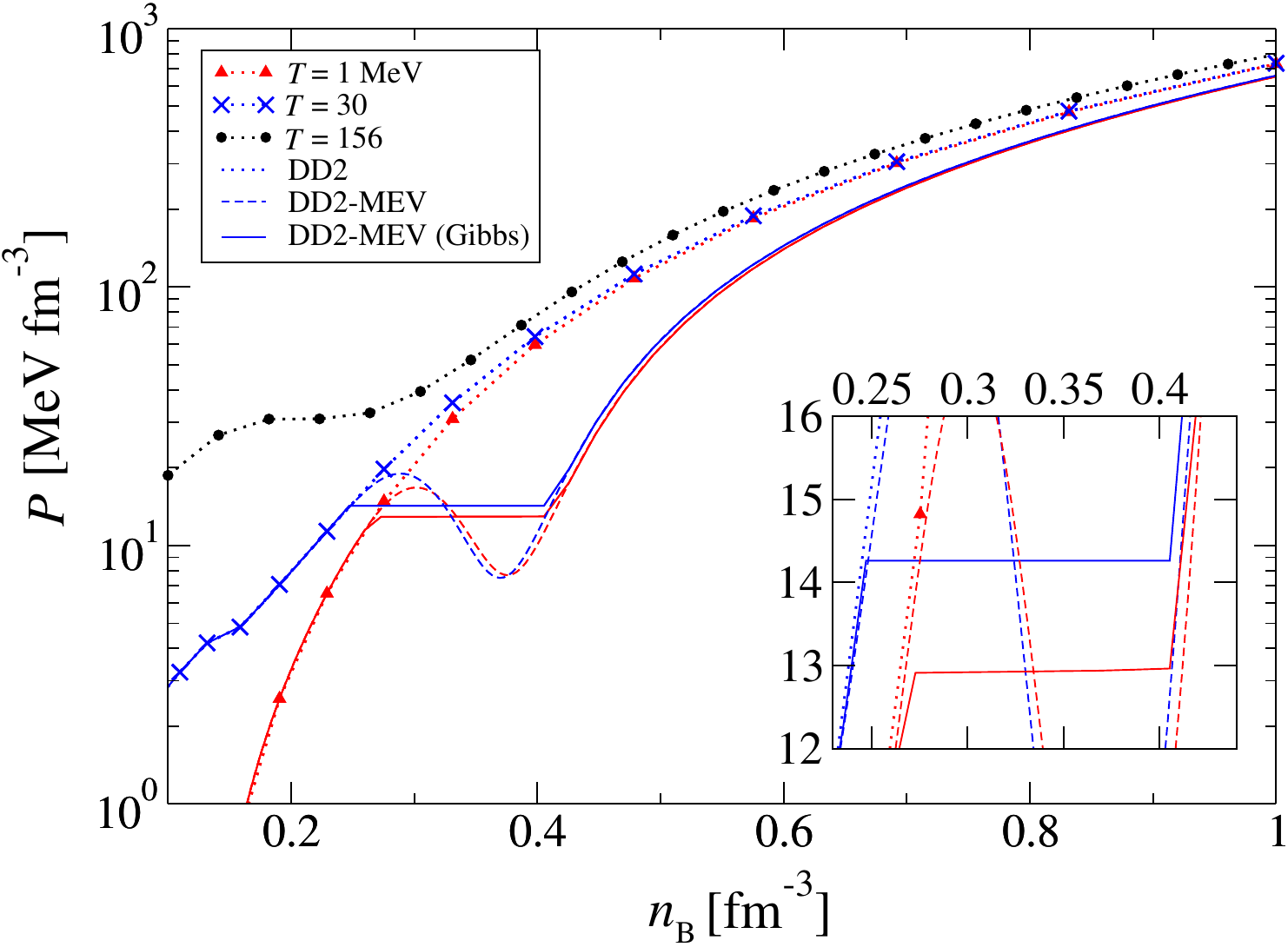}
\label{fig:EOS_b}
}
\hfill
\subfigure[~$T=1$~MeV and varying values of $Y_p$]{
\includegraphics[width=0.49\textwidth]{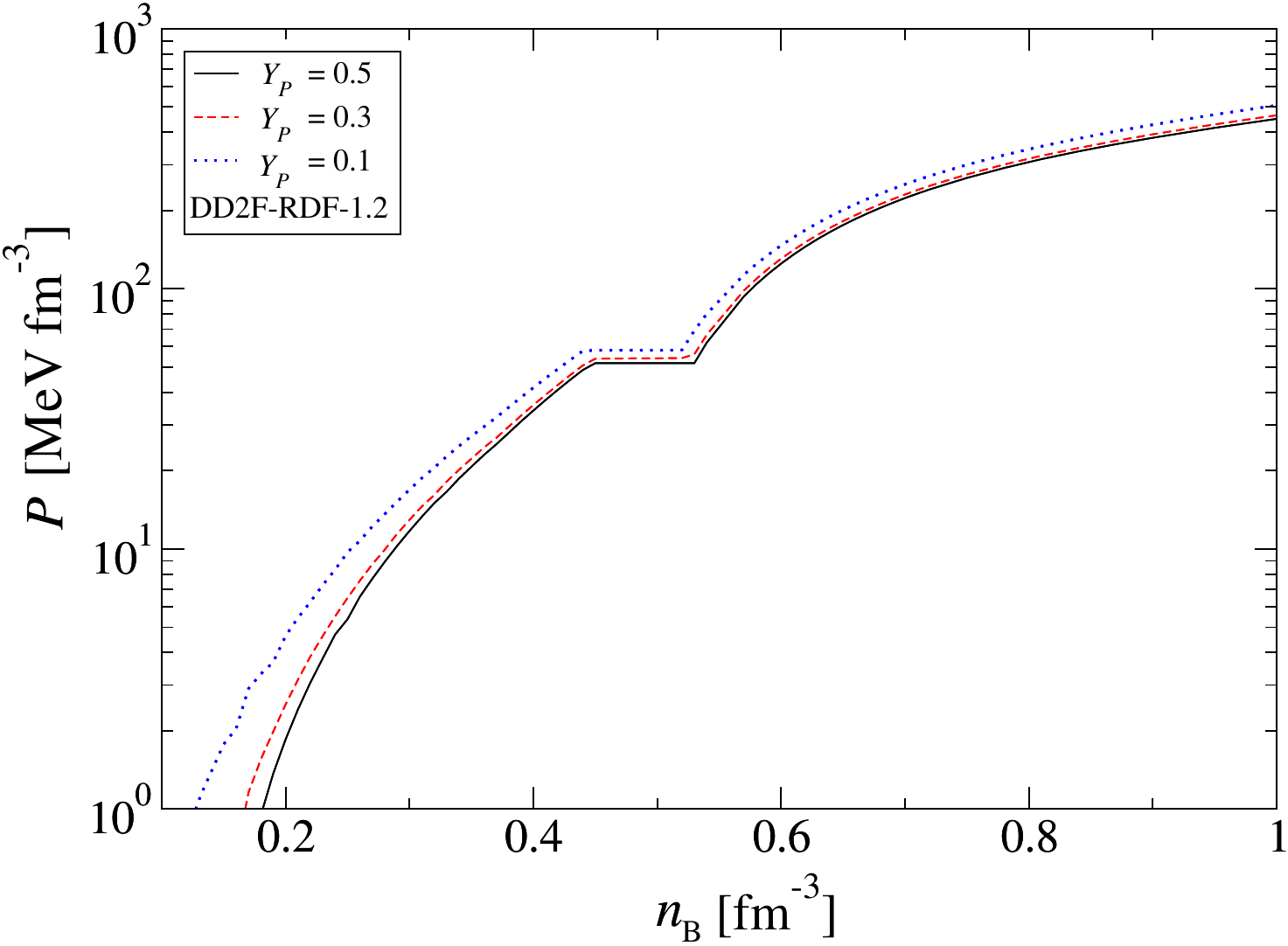}
\label{fig:EOS_c}
}
\hfill
\hspace{-0.75cm}
\subfigure[~$Y_p=0.3$ and two different temperatures]{
\includegraphics[width=0.495\textwidth]{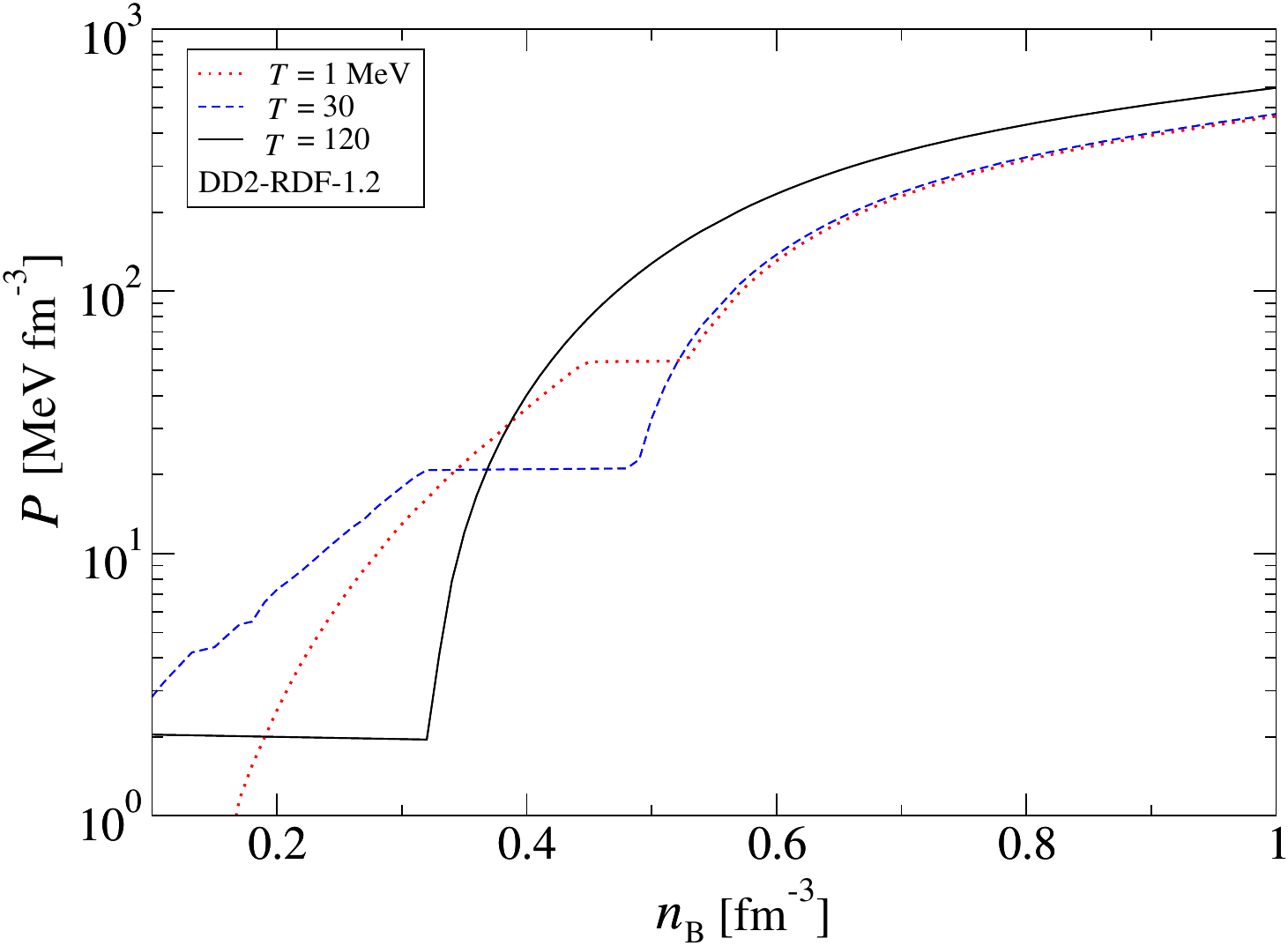}
\label{fig:EOS_d}
}
\caption{EOS comparison, showing the total pressure, $P$, as a function of the baryon density, $n_{\rm B}$, for fixed temperature of $T=1$~MeV and varying proton fractions, $Y_p$ (left panels) as well as for fixed proton fraction of $Y_p=0.3$ and varying temperatures (right panels). 
{\em Top panel:}~DD2-MEV~(Gibbs) with Gibbs phase transition construction (solid lines), DD2-MEV with van der Waals behaviour (dashed lines) and for the DD2 reference EOS (dotted lines).
Note that for the highest temperature of $T=156$~MeV, the first-order phase transition behaviour vanishes, i.e., a the critical point appears. 
{\em Bottom panel:}~Same as top panel but for the DD2F-RDF-1.2 hadron-quark hybrid EOS of Ref.~\cite{Bastian:2021}, featuring a first-order phase transition construction. The highest available temperature is $T=120$~MeV.
}
\label{fig:EOS}
\end{figure}

The MEV variation of DD2, henceforth denoted as DD2-MEV, has a softening effect on the supersaturation density EOS, in comparison to the reference DD2 model. This becomes evident from the choice of parameters \citep[see Table~\ref{tab:eos} and Ref.][]{Typel2018Univ4}. Moreover, due to the algebraic form of the functional $\Phi$, the excluded volume approach features an effective transition from hadronic to quark matter, realized by a change of the effective number of degrees of freedom, via a van der Waals behavior. 
This situation is illustrated in Fig.~\ref{fig:EOS} (dashed lines), showing the pressure $P$ as a function of baryon density $n_{\rm B}$, for low temperature of $T=1$~MeV and varying values of $Y_p$ in Fig.~\ref{fig:EOS_a}, and for fixed $Y_p=0.3$ and varying temperatures in Fig.~\ref{fig:EOS_b}.
For comparison, we also show the corresponding reference DD2 case (dotted lines). 
From this analysis it becomes evident at which conditions, in terms of density, temperature and $Y_p$, the DD2-MEV EOS softens in comparison to the DD2 reference case. 
The slope of the pressure changes drastically and even turning negative, before turning positive again at a higher value of density. This typical van der Waals behaviour, known from the theory of real gases, represents over-critical and under-critical regions, with respect to the potential occurrence of fluctuations. These regions are often discarded as {\em nonphysical} as they might result in imaginary sound velocity or negative entropy. We will return to this point below. 
This is the unique feature of the DD2-MEV model, mimicking a first-order phase transition at supersaturation density, where typically the hadron-quark matter phase transition is assumed, being of first order by design in the common two-phase approach framework~\cite{Glendenning1992,Glendenning:2000csnp.conf.....G}. 

\begin{figure}[t!]
\centering
\subfigure[~][~$T=1$~MeV]{
\includegraphics[width=0.48\textwidth]{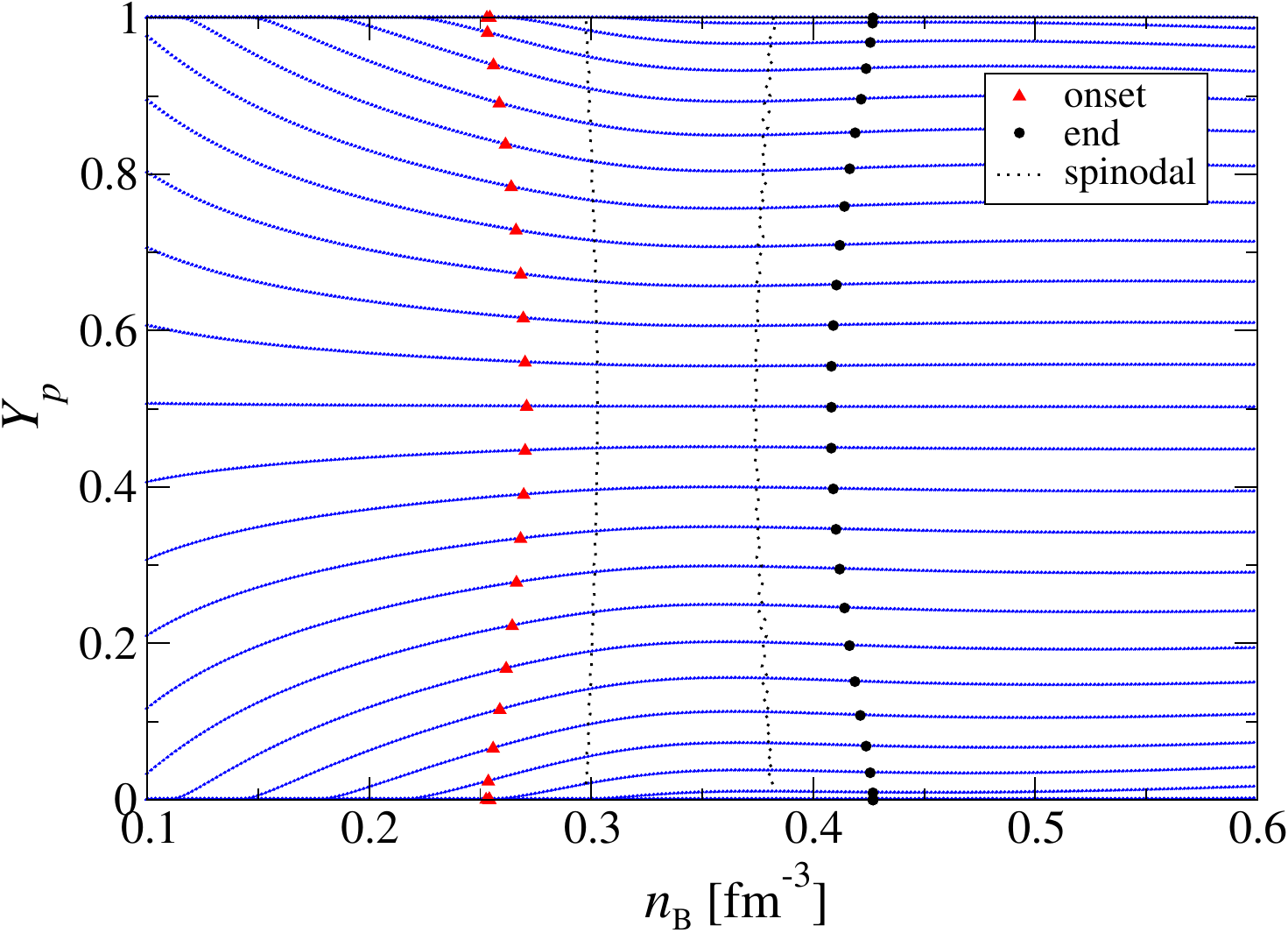}
\label{fig:Phase_const_a}
}
\hspace{-0.5cm}
\subfigure[~][~$T=1$~MeV]{
\includegraphics[width=0.48\textwidth]{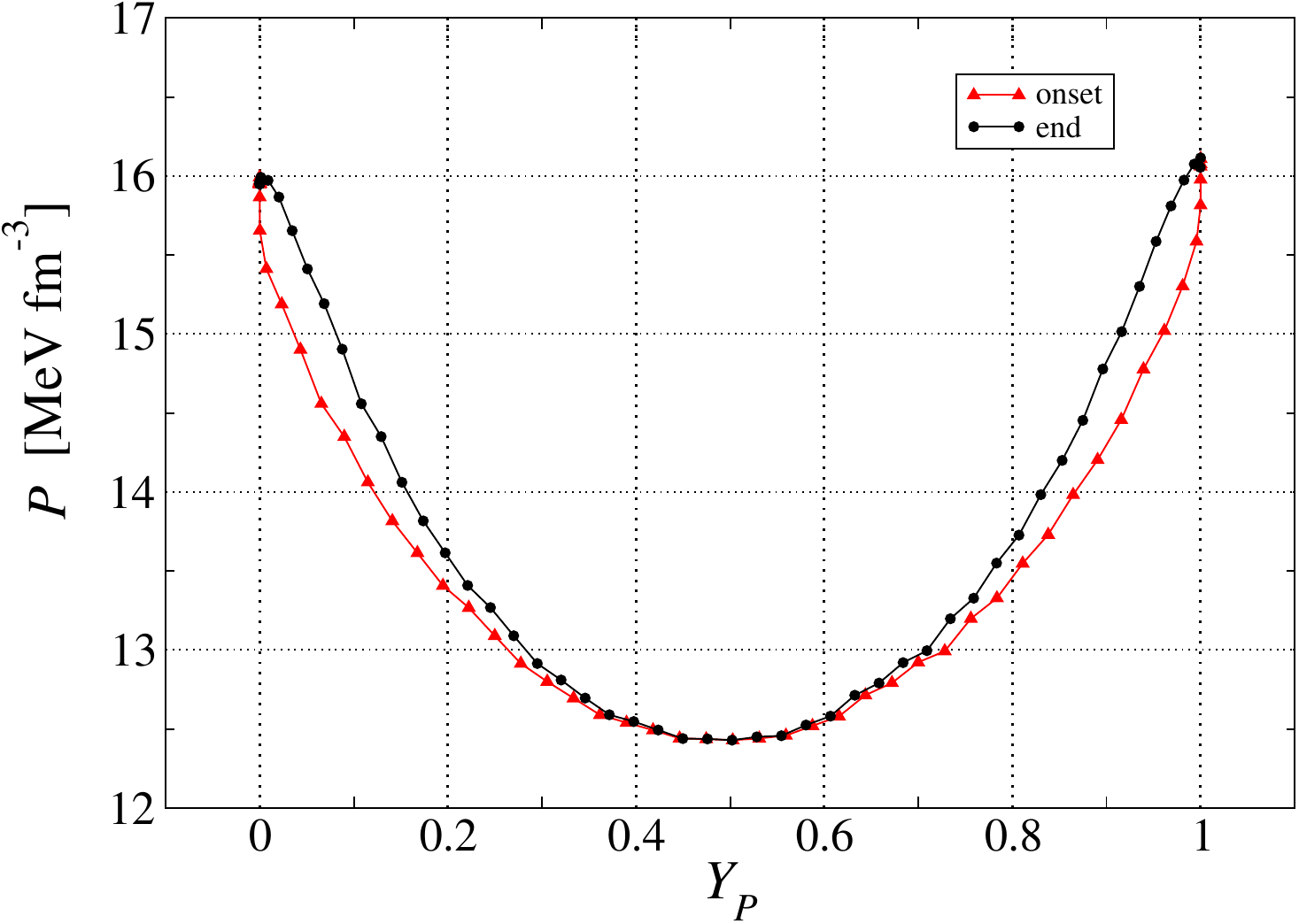}
\label{fig:Phase_const_b}
}
\\
\subfigure[~][~DD2-MEV phase diagram for $Y_p=0.3$]{
\includegraphics[width=0.48\textwidth]{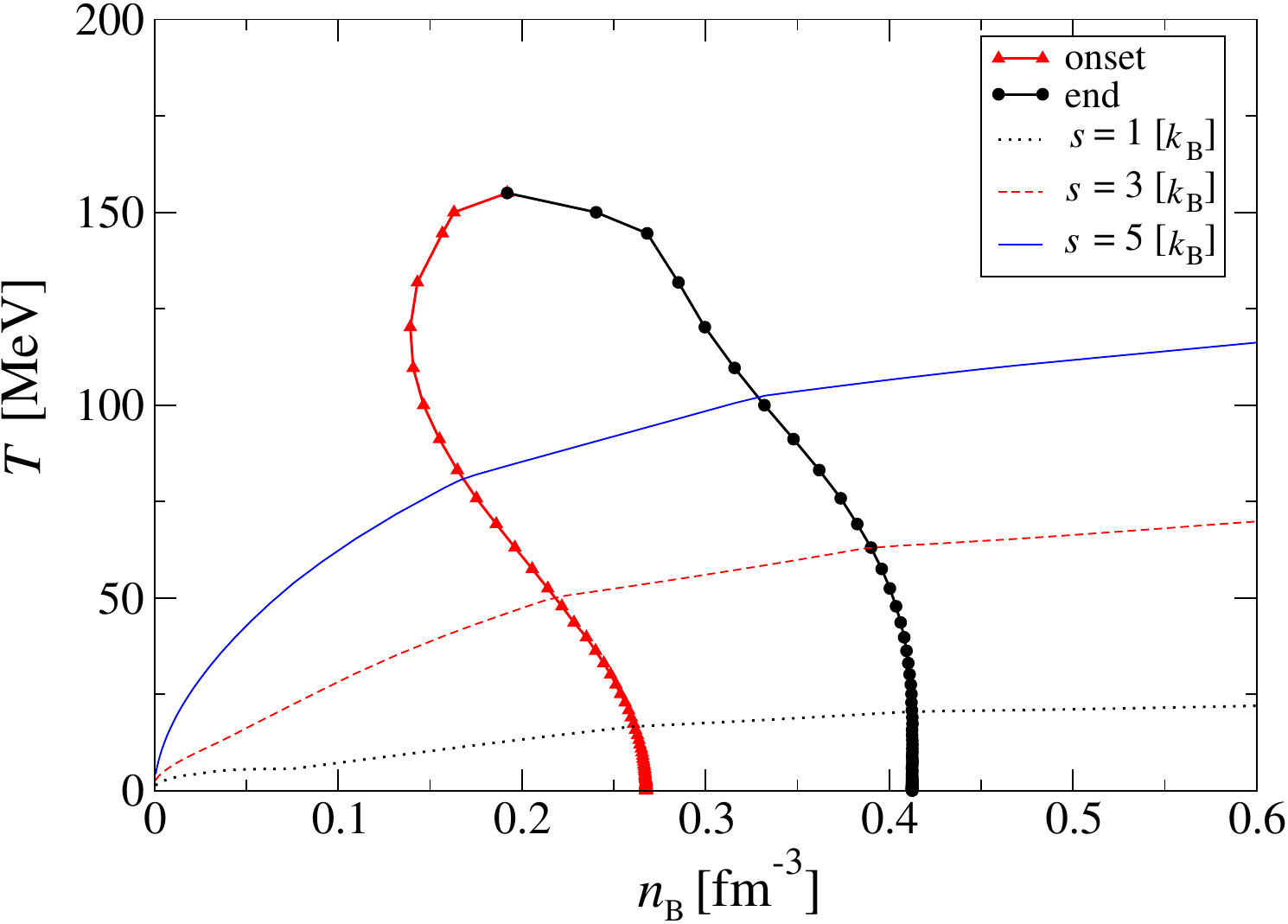}
\label{fig:Phase_const_c}
}
\hspace{-0.5cm}
\subfigure[~][~DD2F-RDF-1.2 phase diagram for $Y_p=0.3$]{
\includegraphics[width=0.48\textwidth]{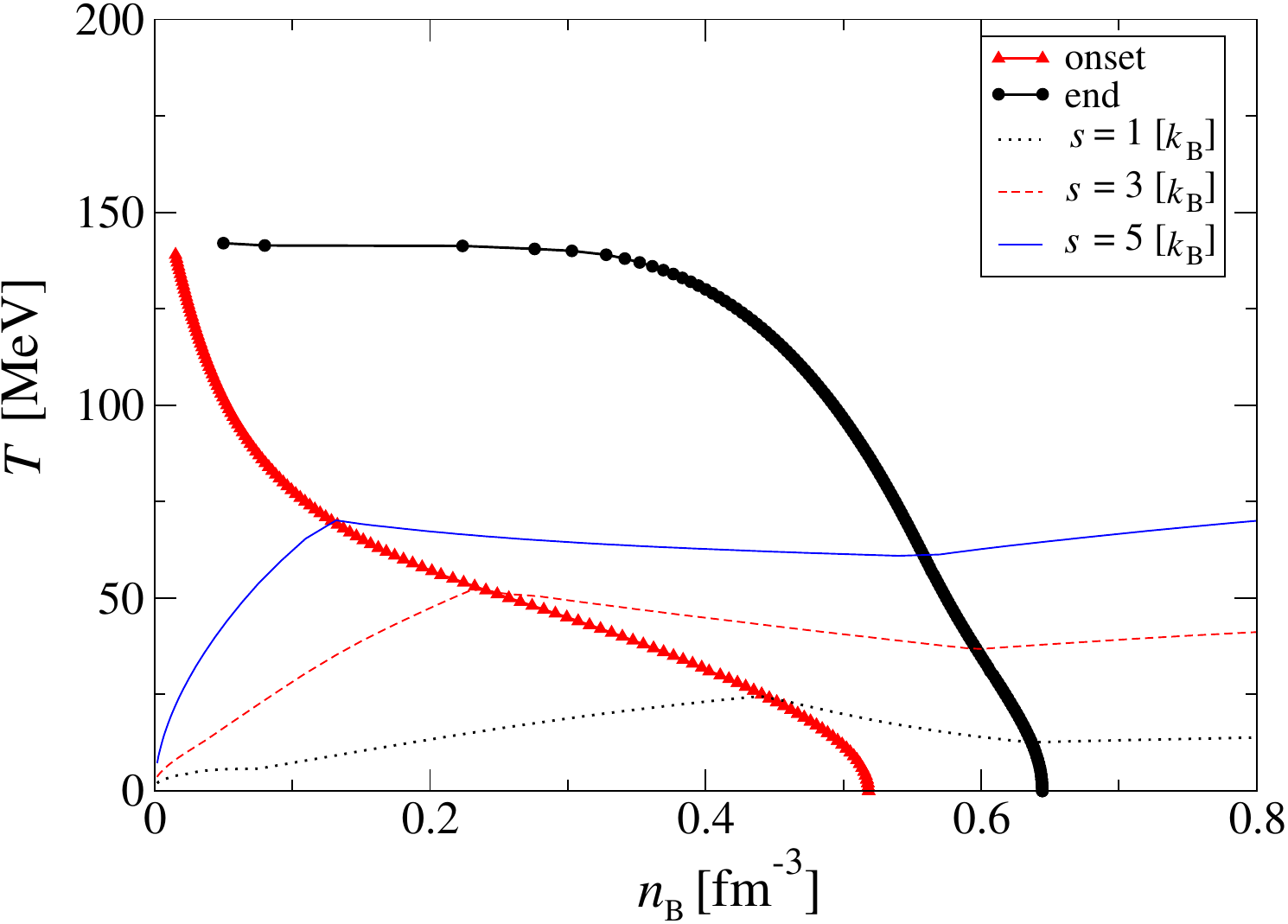}
\label{fig:Phase_const_d}
}
\caption{Binodal region of the DD2-MEV~(Gibbs) EOS at the example of a temperature of $T=1$~MeV. 
Graph~(a):~Curves of equal-charge chemical potential (solid lines) for $Y_p$ as a function of the baryon density, $n_{\rm B}$. Circles indicate the onset (red) and end of the coexisting phase (black) along these lines. The vertical dotted lines represent the spinodal region of instability. 
Graph~(b):~Onset (red) and end pressure curves (black) of the phase transition construction as a function of the proton fraction, $Y_p$, from pure neutron matter ($Y_p=0$) to pure proton matter ($Y_p=1$). 
Graph~(c):~Phase diagram including isentrope curves of constant entropy per particle, $s$, for fixed proton fraction of $Y_p=0.3$, in comparison to the DD2F-RDF-1.2 phase diagram~\cite{Bastian:2021} in graph~(d).}
\label{fig:Phase_const}
\end{figure}

In order to compare the results of DD2-MEV in simulations of core-collapse supernovae (in Sec.~\ref{sec:sim}), with the common approach of a phase transition construction, we employ here the usual Gibbs criteria, i.e., that the intensive thermodynamic quantities $P$, $\mu_{\rm B}$, and $\mu_{Q}$ at given temperature $T$ are identical in the two coexisting phases. In general, it is a non-trivial problem to construct the border 
of the coexistence region, the binodal, if the baryon and charge (or proton) densities are used as independent variable. However, a change
to the baryon density and the charge chemical potential maps the problem to a simple one-dimensional phase construction, see, e.g., Appendix~A in Ref.~\cite{Typel2014EPJA50}. 
This leads to the usual condition of pressure (mechanical) equilibrium and chemical equilibrium~\cite{Hempel2009PhRvD80},
\begin{equation}
P^{\rm I}(T,\mu_Q,n_{\rm B}^{\rm I}) = P^{\rm II}(T,\mu_Q,n_{\rm B}^{\rm II})~.
\end{equation}
Therefore, in order to realize the latter condition, we determine curves of constant charge chemical potential, $\mu_Q$, as illustrated in Fig.~\ref{fig:Phase_const_a}, showing $Y_p$ as a function of the baryon density $n_{\rm B}$. In combination with pressure equilibrium, see Fig.~\ref{fig:Phase_const_b}, we are then able to uniquely determine conditions of phase coexistence, for each temperature, and find the corresponding two values of density, $n_{\rm B}^{\rm I}$ and $n_{\rm B}^{\rm II}$, for the onset of the phase transition and for the end point of the phase transition, denoted as $\rho_{\rm onest}$ and $\rho_{\rm end}$, respectively, in Table~\ref{tab:eos}. 
This strategy enables us to evaluate all thermodynamic quantities in the coexistence region between $\rho_{\rm onest}$ and $\rho_{\rm end}$, such as for the modified free energy density, $\tilde F$, which is a linear function of the volume for constant pressure (see Appendix~\ref{app:Gibbs} for details)

\begin{equation}
\left.\tilde F(\rho)\right\vert_{\rho_{\rm onset}}^{\rho_{\rm end}} = \chi_1(\rho)\,\tilde F_{\rm onset} + (1-\chi_{1}(\rho))\,\tilde F_{\rm end}~, 
\end{equation}
via the definition of a quark volume fraction, in analogy to common two-phase approaches (c.f., Ref.~\cite{Bastian:2021}), as follows,
\begin{equation}
\chi_1(\rho) = \frac{\rho_{\rm end}-\rho}{\rho_{\rm end}-\rho_{\rm onset}}\equiv\chi_{\rm quark}~,
\end{equation}
with $\tilde{F}_{\rm onset/end}=\tilde{F}(\rho_{\rm onset/end})$. 
The corresponding phase boundaries are shown in Fig.~\ref{fig:Phase_const}, for the onset (red circles) and end of the phase transition (black circles).
The resulting EOS with phase transition construction, henceforth denoted as 
DD2-MEV~(Gibbs), is illustrated in Fig.~\ref{fig:EOS} (solid lines). 
We note here that for isospin symmetric matter, i.e. $Y_p=0.5$, the pressure slope in the coexistence region is exactly zero, whereas for constant isospin asymmetry $Y_p<0.5$, there is a finite slope of the pressure depending on the density.

It is important to note, however, that with this phase transition construction based on the DD2-MEV EOS, the density jump between onset and end of the phase transition is significantly weaker than for the class of RDF hybrid EOS reported previously~\cite{Fischer18,Fischer2021EPJA57,Khosravi:2024ApJ964}, which was based on the two-phase approach with the DD2F EOS as hadronic model~\cite{Bastian:2021}. This is illustrated via the corresponding phase diagram of DD2F-RDF-1.2 in Fig.~\ref{fig:Phase_const_d}, in comparison to DD2-MEV~(Gibbs) in Fig.~\ref{fig:Phase_const_c}.
Therefore, see also the qualitatively different behaviors of the pressure for DD2-MEV in Fig.~\ref{fig:EOS_b} and for DD2F-RDF-1.2 in Fig.~\ref{fig:EOS_d}, with increasing temperature, while the $Y_p$ dependence remains weak for both models, as illustrated in Fig.~\ref{fig:EOS_a} for DD2-MEV and in Fig.~\ref{fig:EOS_c} for DD2F-RDF-1.2.
It has severe consequences for the core-collapse supernova explosions, which will be reported and discussed in Sec.~\ref{sec:sim}.

The density of onset (end) of the phase coexistence region is below (above) the density where the pressure reaches its maximum (minimum) as a function of the baryon density for constant temperature as can be seen in Figs.\ref{fig:EOS_a} and \ref{fig:EOS_b}. 
Thus, the binodal encloses the spinodal region where the matter is instable with respect to density fluctuations, indicated by a decrease of the pressure with increasing density. Any density variation will not propagate as a sound wave through the matter, but the perturbation will grow exponentially, leading to a separation of phases.

The construction of the spinodal is, in fact, more complicated since the instability with respect to density fluctuations can be achieved in different ways: by a change in the baryon density, the charge density, or a linear combination of these densities, see, e.g., \cite{PhysRevC.52.2072, PhysRevC.74.024317,PhysRevC.95.055808}. Thus, one has to consider the matrix
\begin{equation}
    M = \left. \left( \begin{array}{cc}
    \frac{\partial^{2} F}{\partial N_{B}^{2}}
    &
    \frac{\partial^{2} F}{\partial N_{B} \partial N_{Q}}
    \\
    \frac{\partial^{2} F}{\partial N_{Q} \partial N_{B}}
    &
     \frac{\partial^{2} F}{\partial N_{Q}^{2}}
    \end{array} \right) \right|_{T,V}
\end{equation}
obtained from second derivatives of the free energy $F(T,V,N_{B},N_{Q})$, assuming an isothermal fluctuation.
Instead of the baryon and charge densities also the neutron and
proton densities could be used as independent variables.
If this matrix has at least one negative
eigenvalue, the matter is instable to a density fluctuation.
For adiabatic fluctuations, the entropy instead of the temperate should be kept constant in the construction of the instability regions.


\section{Simulations of core-collapse supernovae with first-order phase transition}
\label{sec:sim}
In the following, we will briefly introduce our core-collapse supernova model and further discuss the simulation results accordingly with the EOS introduced in Sec.~\ref{sec:eos}.
\subsection{Core-collapse supernova model}
Our core-collapse supernova model---\texttt{AGILE-BOLTZTRAN}---is based on general relativistic neutrino radiation hydrodynamics in spherical symmetry, featuring six-species Boltzmann transport~\cite{Mezzacappa93a,Mezzacappa93b,Mezzacappa93c,Liebendorfer04,Fischer09,Fischer2020PhRvD102}. 
The collision integral employs the standard set of of weak reaction channels \citep[see Table~I in Ref. ][]{Fischer2020PhRvC101}. These include (i) neutrino and anti-neutrino emissivities and absorptivities via to the Urca processes employing the full kinematics and including the (inverse) neutron decay channel~\cite{Fischer2020PhRvC101}, (ii) elastic neutrino scattering on nucleons, commonly referred to as isoenergetic scattering, and coherent scattering on nuclei~\cite{Bruenn85}, (iii) inelastic scattering on electron/positron~\cite{Bruenn85,Mezzacappa93c} and muon/antimuon~\cite{Fischer2020PhRvD102}, and (iv) neutrino pair production and absorption processes. The latter include the classical electron-positron annihilation reactions~\cite{Bruenn85} as well as nucleon-nucleon bremsstrahlung~\cite{hannestad98} including leading order medium contributions to the nucleon-nucleon-pion vertex~\cite{Fischer2016A&A593A} and the annihilation of electron neutrino pairs to muon/tau neutrino pairs~\cite{Buras06a,Fischer09}.
For the present investigation, muons and associated muonic weak reactions are being neglected, i.e. $\nu_{\mu}\equiv\nu_{\tau}$ and $\bar\nu_{\mu}\equiv\bar\nu_{\tau}$. 

\texttt{AGILE-BOLTZTRAN} has a flexible EOS module~\cite{Hempel12}. 
The nuclear EOS are based on the relativistic mean field (RMF) models of Ref.~\cite{Hempel2010NuPhA837}, of which we select the stiff DD2 model, featuring density-dependent meson-nucleon couplings~\cite{Typel:2009sy}.
Heavy and light nuclei, which are abundant at subsaturation density and temperatures below $T\leq 15$~MeV, are treated via the modified nuclear statistical equilibrium (NSE) approach~\cite{Hempel2010NuPhA837}, including Coulomb contributions. 
At temperatures below $T\leq 0.5$~MeV, \texttt{AGILE-BOLTZTRAN} assumes a transition to the silicon-sulfur gas, resembling closely the location of the silicon-sulfur shells above the stellar iron core for most iron-core collapse supernova progenitors~\cite{Woosley:2002zz}. 
The electron, positron and photon EOS is based on Ref.~\cite{Timmes1999}, including Coulomb contributions.

We also employ the class of RDF hybrid EOS featuring a first-order phase transition construction from normal nuclear matter to deconfined quark matter~\cite{Kaltenborn17,Bastian:2021}. This is essential in order to explore supernova explosions that are triggered by a sufficiently strong first-order phase transition~\cite{Fischer18,Fischer2021EPJA57,Khosravi:2024ApJ964}. 
This class of EOS is characterized by a density jump at supersaturation densities from the onset conditions for the appearance of quark matter to pure the quark matter phase, due to a phase transition construction based on the Gibbs approach of phase coexistence in the presence of more than one conserved charges~\cite{Hempel2009PhRvD80,Bastian:2021}. 
The presence of a region of thermodynamic and hydrodynamic instability, located in between the two stable hadronic and quark matter phases, is key for the associated supernova explosion mechanism. This aspect will be further discussed below.

In addition to these hadronic and hybrid model EOS, we implement the newly developed DD2-MEV EOS (see Sec.~\ref{sec:eos}). In the following we distinguish two versions, one with the unmodified van der Waals behavior, featuring over- and under-critical regions, e.g., negative pressure slope, which has a continuous behavior across the phase transition region, and a second model in which we implement a Gibbs construction for the phase transition. The latter, henceforth denoted as DD2-MEV (Gibbs), removes the over- and under-critical region.

\subsection{Supernova simulations with the class of DD2-MEV EOS}
Simulations are launched from the stellar progenitor with a ZAMS mass of 40~M$_\odot$, from the stellar evolution series of Ref.~\cite{Rauscher:2002}, henceforth denoted as \texttt{s40a28} in accordance with Ref.~\cite{Rauscher:2002}. 
Massive progenitors in the zero-age main sequence mass range of 25--75~M$_\odot$ have been explored recently as candidates that feature core-collapse supernova explosions driven by sufficiently strong first-order phase transition from normal nuclear (in general hadronic) matter to deconfined quark matter (c.f. Refs.~\cite{Fischer18,Zha20,Fischer:2021jfm,JakobusMueller2022,Kuroda2022ApJ924_QCD,Khosravi:2024ApJ964}, and references therein).
In those studies, Gibbs and Maxwell criteria for phase equilibrium have been employed for the first-order phase transition construction, which has been argued to be a key input for the success of the underlying explosion mechanism. 
Following the same strategy, we first discuss results here that are obtained with the DD2-MEV~(Gibbs) EOS, which was introduced in Sec.~\ref{sec:eos}.
We emphasize here that DD2-MEV~(Gibbs) features a significantly less pronounced density jump between the two stable hadronic and quark matter phases, in comparison to the previously studied RDF models, in particular those which have been found to be more favorable for CCSN explosions (see therefore the EOS in Fig.~\ref{fig:EOS} and the phase diagrams in Figs.~\ref{fig:Phase_const_c}, for DD2-MEV, and \ref{fig:Phase_const_d}, for DD2F-RDF-1.2).
As a consequence, we find that with DD2-MEV~(Gibbs) CCSN explosions can no longer be obtained. 

\begin{figure}[t!]
\centering
\subfigure[~DD2-MEV~(Gibbs)]{
\includegraphics[width=0.397\textwidth]{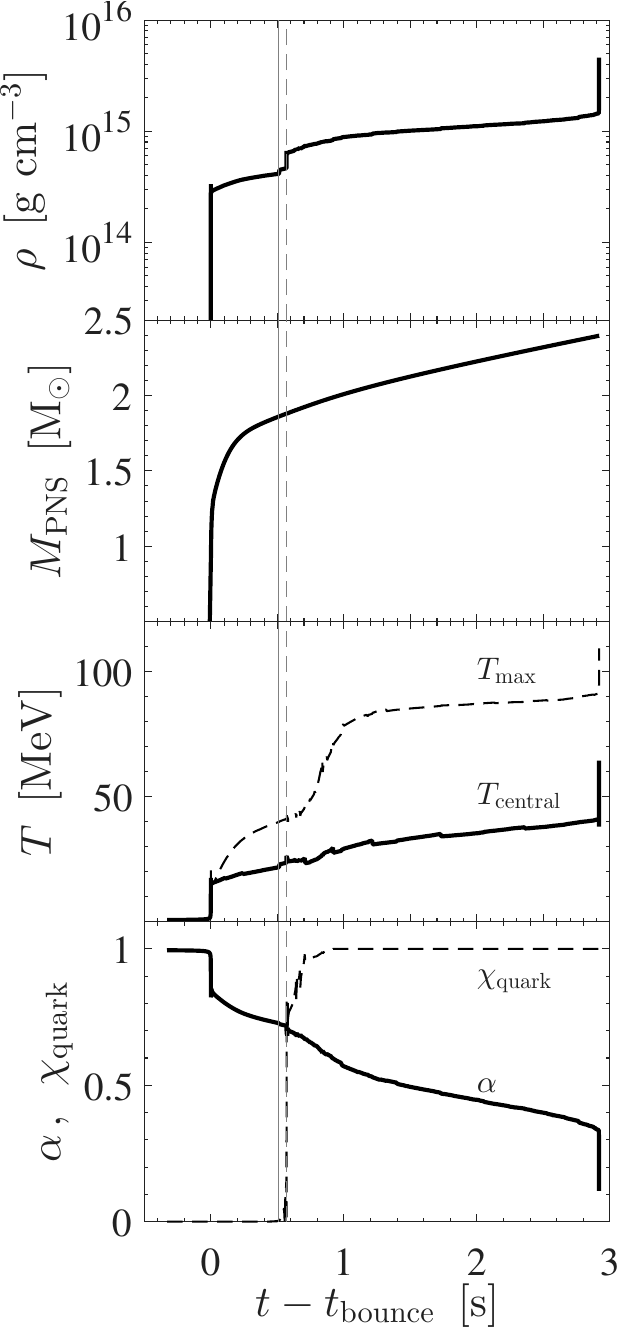}
\label{fig:SN-central_a}
}
\subfigure[~DD2-MEV]{
\includegraphics[width=0.39\textwidth]{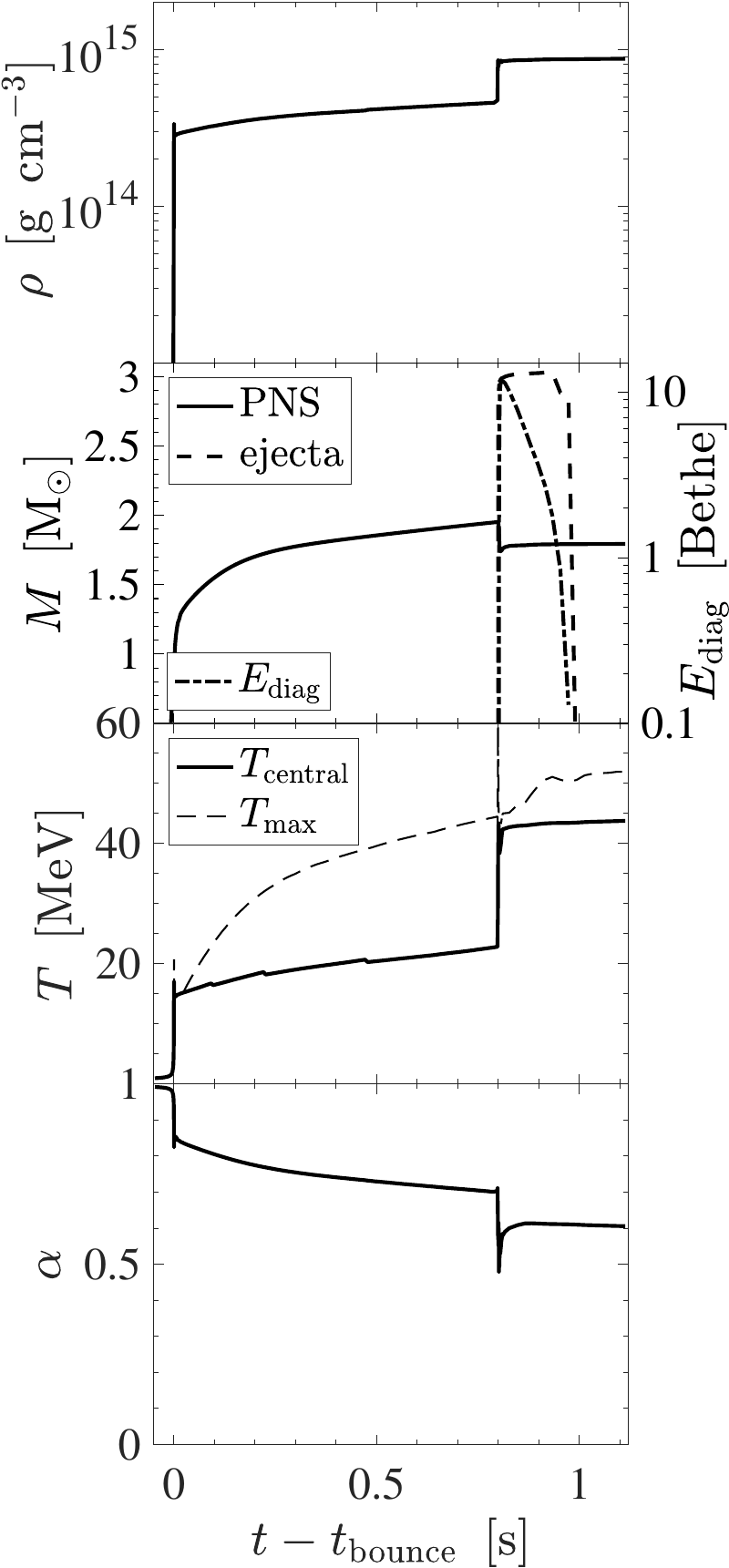}
\label{fig:SN-central_b}
}
\caption{Post bounce evolution of selected quantities for the runs with DD2-MEV~(Gibbs) EOS in graph~(a) and DD2-MEV with van der Waals behaviour in graph~(b), showing (from top to bottom), central restmass density $\rho$, enclosed PNS mass $M_{\rm PNS}$, central and peak temperatures, $T_{\rm central}$ and $T_{\rm max}$, respectively, and the central lapse function $\alpha$. 
The gray solid and dashed vertical lines in graph~(a) mark times for the onset of the phase transition and the time of the PNS contraction, respectively, and the bottom panel in graph~(a) contains the central quark fraction $\chi_{\rm quark}$ (see text for details).
In addition, the DD2-MEV simulation shows the ejecta mass (dashed lines) and the explosion energy estimate, denoted as $E_{\rm diag}$ (dash-dotted line).
\label{fig:SN-central}
}
\end{figure}
\begin{figure}[t!]
\centering
\subfigure[~DD2-MEV~(Gibbs)]{
\includegraphics[width=0.4\textwidth]{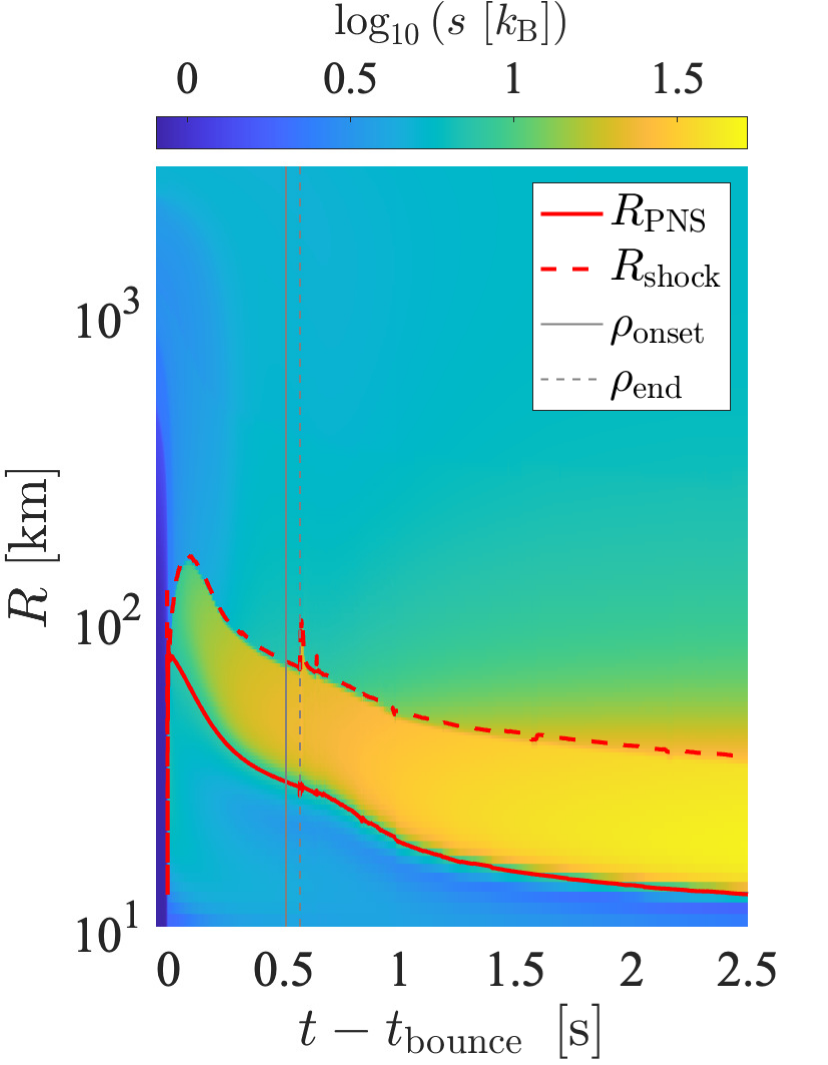}
\label{fig:SN-evol_a}
}
\subfigure[~DD2-MEV]{
\includegraphics[width=0.4\textwidth]{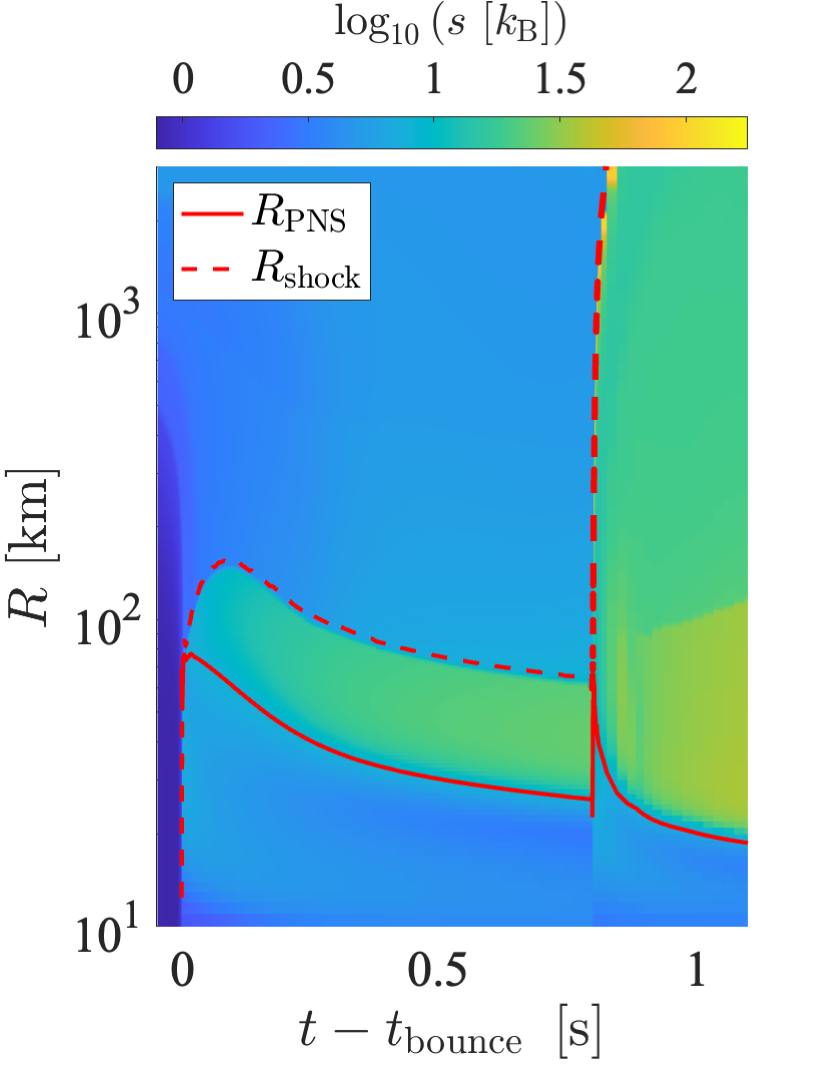}
\label{fig:SN-evol_b}
}
\caption{Post bounce evolution the radial evolution of the simulation domain up to few $R=10^3$~km, with entropy per baryon, $s$, color coded in log-scale, comparing the DD2-MEV~(Gibbs) run (left panel) and DD2-MEV (right panel).
The locations of the PNS radius $R_{\rm PNS}$ and shock $R_{\rm shock}$ are marked by solid and dashed red lines, respectively. 
For the DD2-MEV~(Gibbs) simulation, we mark the time for reaching the onset density for the phase transition at $t\simeq 0.51$~s post bounce, denoted as $\rho_{\rm onset}$  (vertical solid gray line), and the pure quark matter phase at $t\simeq 0.57$~s post bounce, denoted as $\rho_{\rm final}$ (vertical dashed gray line).
\label{fig:SN-evol}
}
\end{figure}

The subsequent post-bounce evolution for the CCSN simulation with DD2-MEV~(Gibbs) is illustrated in Fig.~\ref{fig:SN-central_a}, showing the post-bounce evolution of selected quantities, from top to bottom central restmass density $\rho$, enclosed PNS mass $M_{\rm PNS}$, central and maximum temperatures, $T_{\rm central}$ and $T_{\rm max}$, respectively, and central lapse function $\alpha$ and quark matter volume fraction $\chi_{\rm quark}$.
The central region of the PNS reaching sufficiently high enough densities for the phase transition to occur (top panel), after about 0.5~s post bounce. We note here that the conditions, c.f. restmass density and temperature in Fig.~\ref{fig:SN-central} are comparable to those previously assumed~\cite{Bastian:2021}.
However, unlike what was reported in previous supernova simulations, featuring a first-order phase transition, here the PNS fails to collapse supersonically. Instead, we find a mild hydrodynamical PNS structure reconfiguration, as the central regions undergo the phase transition. 
The PNS contracts mildly and then expands for a short timescale of few tens of milliseconds (see the red solid line in Fig.~\ref{fig:SN-evol_a}), and the central density increases from  about $\rho\simeq 5.5\times 10^{14}$~g~cm$^{-3}$ to $\rho\simeq 8\times 10^{14}$~g~cm$^{-3}$ (top panel in Fig.~\ref{fig:SN-central_a}), on the same timescale. 
As a hydrodynamics feedback response to the PNS contraction, the bounce shock expands (see the red dashed line in Fig.~\ref{fig:SN-evol_a}), from about $R_{\rm shock}=69$~km before the phase transition to $R_{\rm shock}=100$~km, before retracting again to about $R_{\rm shock}=65$~km after the phase transition.
Immediately after the phase transition, the central quark volume fraction (bottom panel in Fig.~\ref{fig:SN-central_a}) reached values of about $\chi_{\rm quark}\simeq 0.8$, i.e. pure quark matter has not been obtained at that evolution stage. 
Only during the later evolution, on the order of several hundreds of milliseconds, the PNS contracts due to the continuous mass accretion in the absence of an explosion, and central density and temperature increase, such that pure quark matter is obtained when $\chi_{\rm quark}=1$ at about 1~s post bounce. 
Note that the timescale for the rise of central density and temperature is increased after the phase transition, since the EOS is softer in the pure quark and in the quark-hadron mixed phases than in the hadronic phase, reaching central densities in excess of $\rho\simeq 10^{15}$~g~cm$^{-3}$ and $T_{\rm max}\simeq 80$~MeV, at about 1.5~s post bounce. 

The later evolution of the DD2-MEV~(Gibbs) run is qualitatively comparable to the standard failed core-collapse supernova explosion phenomenology, i.e. mass accretion will continue onto the PNS until the enclosed PNS mass exceeds the maximum mass, given in turn by the EOS. 
After this, stable solutions of the radiation hydrodynamics equations in co-moving coordinates can be obtained no more and the simulations are stopped.
Recently, supernova simulations could extend beyond this point by the implementation of a horizon finder and the removal of the singular domains from the simulations~\cite{Rahman2022MNRAS512,Kuroda2023MNRAS526_beyond-failed-SNe,OConnor2026arXiv260501405E}.
In the further post-bounce evolution after the phase transition, the DD2-MEV~(Gibbs) simulation proceeds towards the black hole formation at a  post bounce time of $t-t_{\rm bounce}=2.918$~s.

\begin{figure}[t!]
\centering
\subfigure[~PNS collapse and instability]{ 
\includegraphics[width=0.975\textwidth]{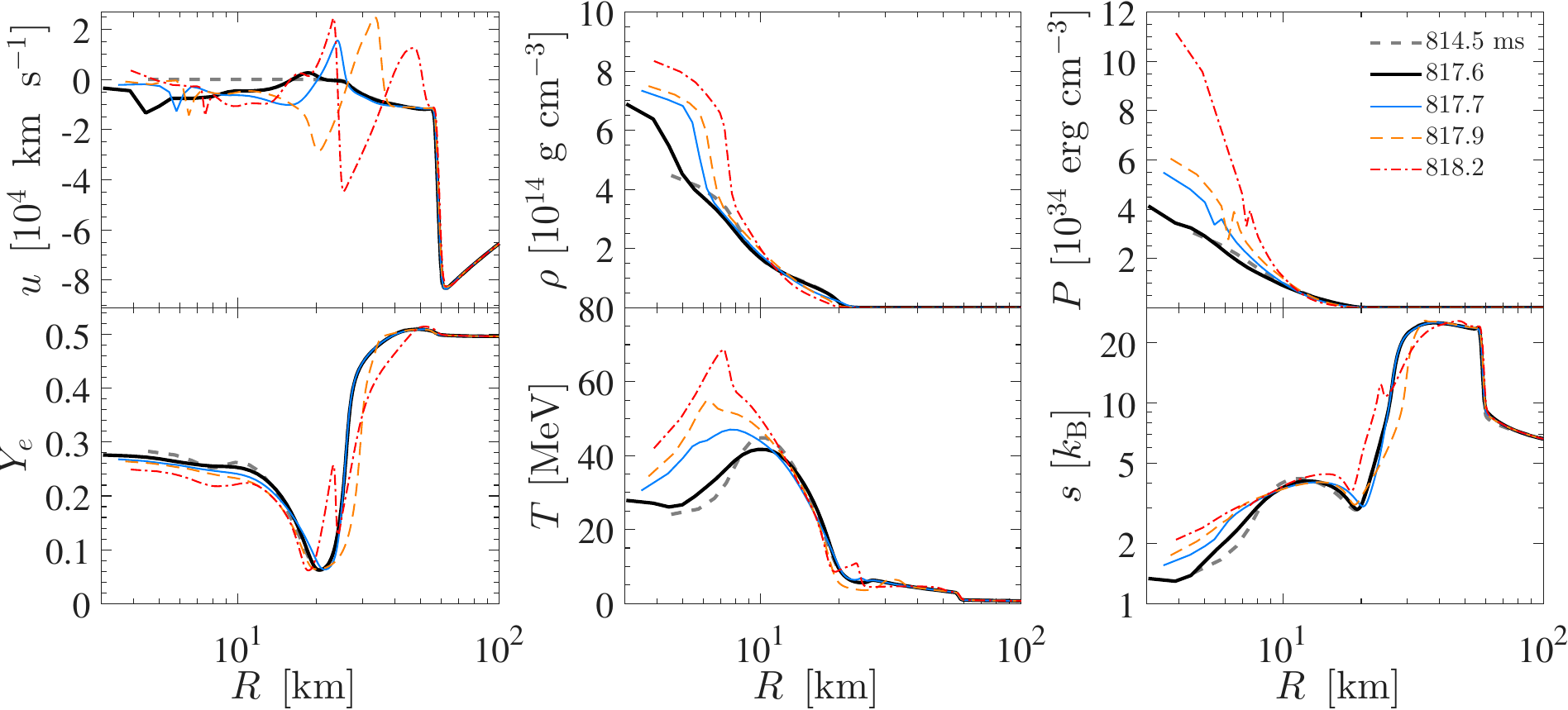}
\label{fig:SN-profiles_a}
}
\\
\subfigure[~Explosion onset]{
\includegraphics[width=0.975\textwidth]{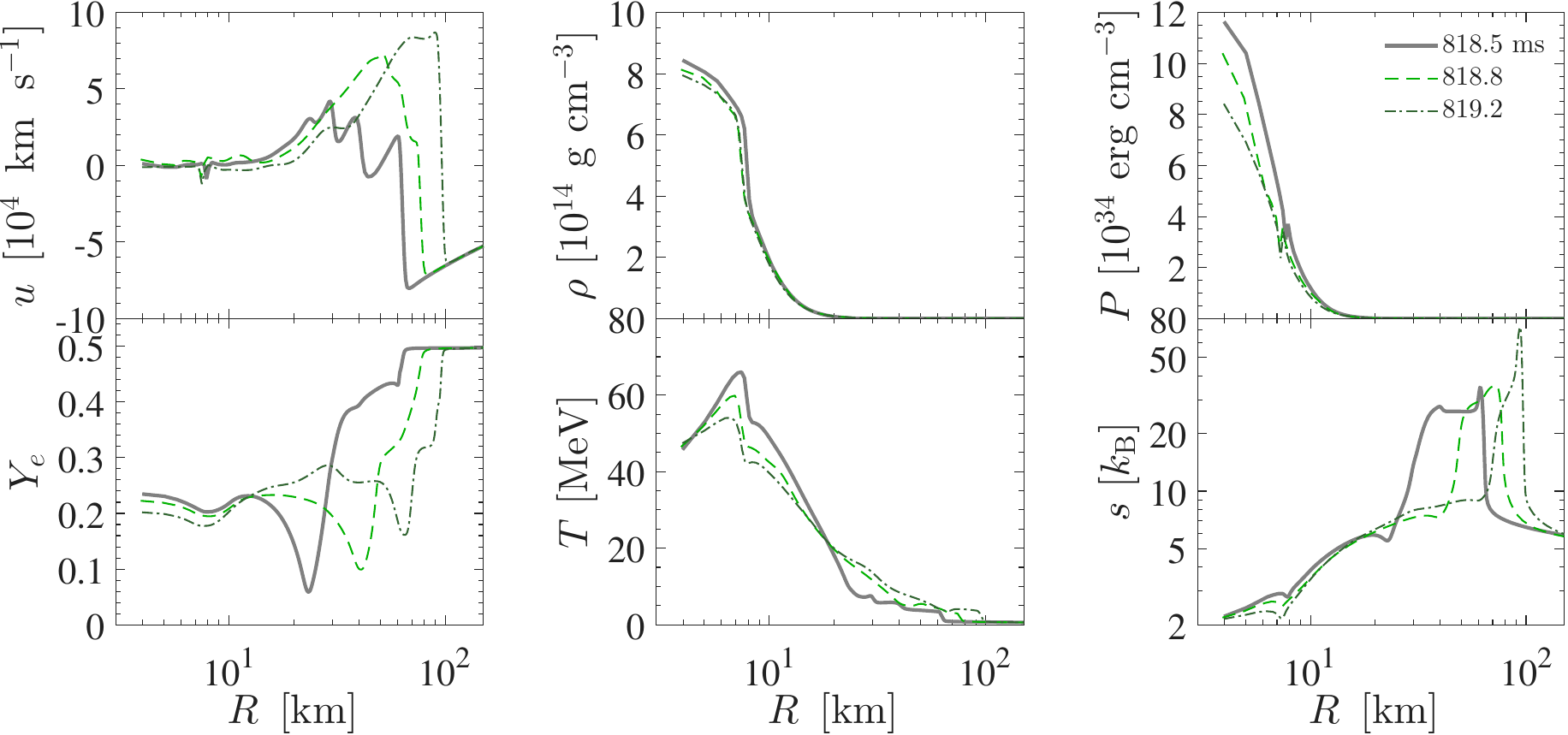}
\label{fig:SN-profiles_b}
}
\caption{Radial profiles of selected quantities during the early evolution after the phase transition for the simulations with the DD2-MEV EOS, featuring van der Waals behaviour for the first-order phase transition, showing radial velocity $u$, restmass density $\rho$, pressure $P$, electron fraction $Y_e$, temperature $T$ and entropy per baryon $s$. 
\label{fig:SN-profiles}
}
\end{figure}

The situation with DD2-MEV~(Gibbs) appears qualitatively different for the simulation with the DD2-MEV EOS, i.e. with van der Waals behavior for the first-order phase transition.
While for the case of DD2-MEV~(Gibbs) with Gibbs phase transition construction. 
An explosion is observed in the case of DD2-MEV. We note here that both DD2-MEV and DD2-MEV~(Gibbs) feature the very same nuclear matter and bulk matter properties. the only difference arises due to the presence of a phase transition construction for DD2-MEV~(Gibbs) whereas DD2-MEV keeps the van der Waals behavior.
Furthermore, simulation results reported previously of CCSN explosions triggered by a sufficiently strong first-order phase transition~\cite{Sagert09,Fischer18,Zha20,Fischer:2020,Fischer:2021jfm,Kuroda2022ApJ924_QCD,Khosravi:2024ApJ964} are being reproduced here qualitatively.

The post-bounce evolution of the same selected quantities is illustrated in Fig.~\ref{fig:SN-central_b}.
It becomes evident that the central density rises sharply as a consequence of the phase transition, indicating a qualitatively different behavior than in the case with phase transition construction (compare with Fig.~\ref{fig:SN-central_a}). 
In fact, the PNS collapse results in supersonic velocities, leading to the formation of a strong second hydrodynamic shock wave. 
However, unlike in the cases reported previously, in which the second shock forms at the phase boundary between infalling hadronic matter and deconfined quark matter, here initially we find no strong shock wave formation, only a hydrodynamics perturbation wave, as is illustrated in the upper left panel of Fig.~\ref{fig:SN-profiles_a}. 
This perturbation grows while moving outwards in time, see $t=t_{\rm bounce}=817.6$~ms--818.2~ms ($t-t_{\rm bounce}=814.5$~ms represents the conditions just before the phase transition occurs), during which matter is still falling onto the bounce shock, located at around 60~km, resulting in the continuous increase of the central restmass density, temperature and hence pressure as well as entropy per baryon, while the central electron fraction $Y_e$\footnote{~In the absence of other leptonic charges, $Y_e\equiv Y_p$.} remains unchanged (see Fig.~\ref{fig:SN-profiles_a}). 

Note the negative pressure slope (upper right panels in Fig.~\ref{fig:SN-profiles}), between $P=3$--$4\times 10^{34}$~erg~cm$^{-3}$, which is due to the van der Walls behavior of the first-order phase transition, implemented via the excluded volume mechanism, as was discussed in Sec.~\ref{sec:eos}. The associated over- and under-compressed regions imply imaginary sound velocity, $cs$, in the presence of thermodynamic and hydrodynamic instability, i.e. $cs^2<0$ with $cs^2=\left(\partial P/\partial\epsilon\right)_{s={\rm const.}}$\footnote{~$cs$ has units of the speed of light.} and since the energy density, $\epsilon$, is a smooth function across the phase transition (for a review about the topic of hydrodynamical instabilities in classical fluids, c.f. Ref.~\cite{ChomazColonnaRandrup:2004PhR389} and references therein).
However, since in the Lagrangian formulation of the neutrino radiation hydrodynamics equations of {\tt AGILE-BOLTZTRAN} the sound speed does appear explicitly. Nevertheless, in order to ensure numerical stability, we ignore imaginary sound speed and apply instead $cs\equiv 0$\footnote{~For computational purposes, we select $10^{-10}$ in units of the speed of light as minimum speed of sound.}, also for the gravitational wave mode analysis below in Sec.~\ref{sec:modes}. 

Once the sub-sonic hydrodynamic perturbation wave reaches the standing bounce shock, matter velocities increase (upper left panels in Figs.~\ref{fig:SN-profiles_a} and \ref{fig:SN-profiles_b}) to supersonic velocities, along the continuously decreasing density gradient of the PNS surface. At the bounce shock, this perturbation front turns into a shock wave, with matter velocities reaching on the order of 30\% of the speed of light (Fig.~\ref{fig:SN-profiles_b}).
The still infalling matter from the outer layers of the stellar core is shock heated, reaching matter temperatures exceeding $T=70$~MeV (see also the evolution of the central fluid element in Fig.~\ref{fig:SN-evol_b} with central and maximal temperatures reached, as well as with the central lapse function decreasing rapidly, from $\alpha\simeq 0.7$ to $\alpha\simeq 0.5$). 

There are two major differences in the evolutionary paths towards explosion of the DD2-MEV model, in comparison to the previously reported QCD-driven supernova explosions, namely, previous bag~\cite{Sagert09,Fischer11} and RDF models~\cite{Fischer17,Fischer2021EPJA57,Bastian:2021,Khosravi:2024ApJ964} for quark matter:
\begin{enumerate}
\item The latter quark-hadron hybrid models feature generally lower temperatures in the quark matter phase, compared to the hadronic EOS, for a given isentrope, i.e. the temperature tends to decrease during the phase transition for matter having low entropy, on the order of few $k_{\rm B}$ per baryon. 
For the DD2-MEV model, we note the opposite behavior, which has already been pointed out in Sec.~\ref{sec:eos}. 
As a consequence, even with strong compression due to a supersonic collapse, as was reported previously, the central temperature increases rapidly during the phase transition, from $T\simeq 30(40)$~MeV to $T\simeq 40(70)$~MeV, at the center(peak), as can be seen in Fig.~\ref{fig:SN-profiles_a}.
In the previously reported QCD driven explosion models, an initial temperature decrease was observed, followed by a rapid supersonic compression, during which too the central temperature started to increase where due to the presence of a strong hydrodynamic shock wave early on, the entropy was not constant anymore.
\item Second, it takes substantially longer between PNS collapse, with the formation of the stagnation front, and turning into an expanding shock wave. This is related to the formation of the shock wave only when reaching the bounce shock, which is qualitatively different from the previously reported evolutions of MIT-bag models~\cite{Sagert09,Fischer11} as well as RDF models~\cite{Kuroda2022ApJ924_QCD,Khosravi:2024ApJ964}. 
The propagation of the initially formed sub-sonic stagnation front lasts several milliseconds. 
The extended delay between PNS collapse and shock acceleration is also reflected in the somewhat slower PNS radius expansion and contraction, as illustrated in Fig.~\ref{fig:SN-evol_b} (red solid line), which lasts several tens of milliseconds. This aspect has important consequences for the neutrino signal, which will be discussed below.
\end{enumerate}

As a direct consequence, the rise of the central density is somewhat less sudden than previously reported and, related, the mass ejected from the PNS is substantially larger, on the order of 0.2~M$_\odot$ (see Fig.~\ref{fig:SN-central_b}). Note that mass ejection from the newly formed PNS with quark matter core was previously reported on the order of few $10^{-2}$~M$_\odot$ \cite{Fischer:2020}. Hence, the diagnostic explosion energy estimate, denoted as $E_{\rm diag}$ \citep[for a definition, c.f. expressions~(8)--(11) in Ref.~][]{Fischer10}, rises sharply to values in excess of $E_{\rm diag}\simeq 10$~Bethe\footnote{$1~{\rm Bethe}\equiv 10^{51}~{\rm erg}$.} , which is one order of magnitude larger than previously reported (see Fig.~\ref{fig:SN-central_b}). The estimated total ejecta mass (see Fig.~\ref{fig:SN-central_b}), defined via the material located outside the mass cut, reaches values of about $M_{\rm ejecta}\simeq 3$~M$_\odot$. 
However, during the further, the dynamical mass ejection suffers from a large fall back. The ejecta mass drops below $M_{\rm ejecta}\simeq 1$~M$_\odot$ and the diagnostic explosion energy estimate drops accordingly towards $E_{\rm diag}\simeq 0.1$~Bethe. 

\section{Neutrino signal}
\label{sec:neutrino}
\begin{figure}[t!]
\centering
\subfigure[~DD2-MEV~(Gibbs)]{
\includegraphics[width=0.4\textwidth]{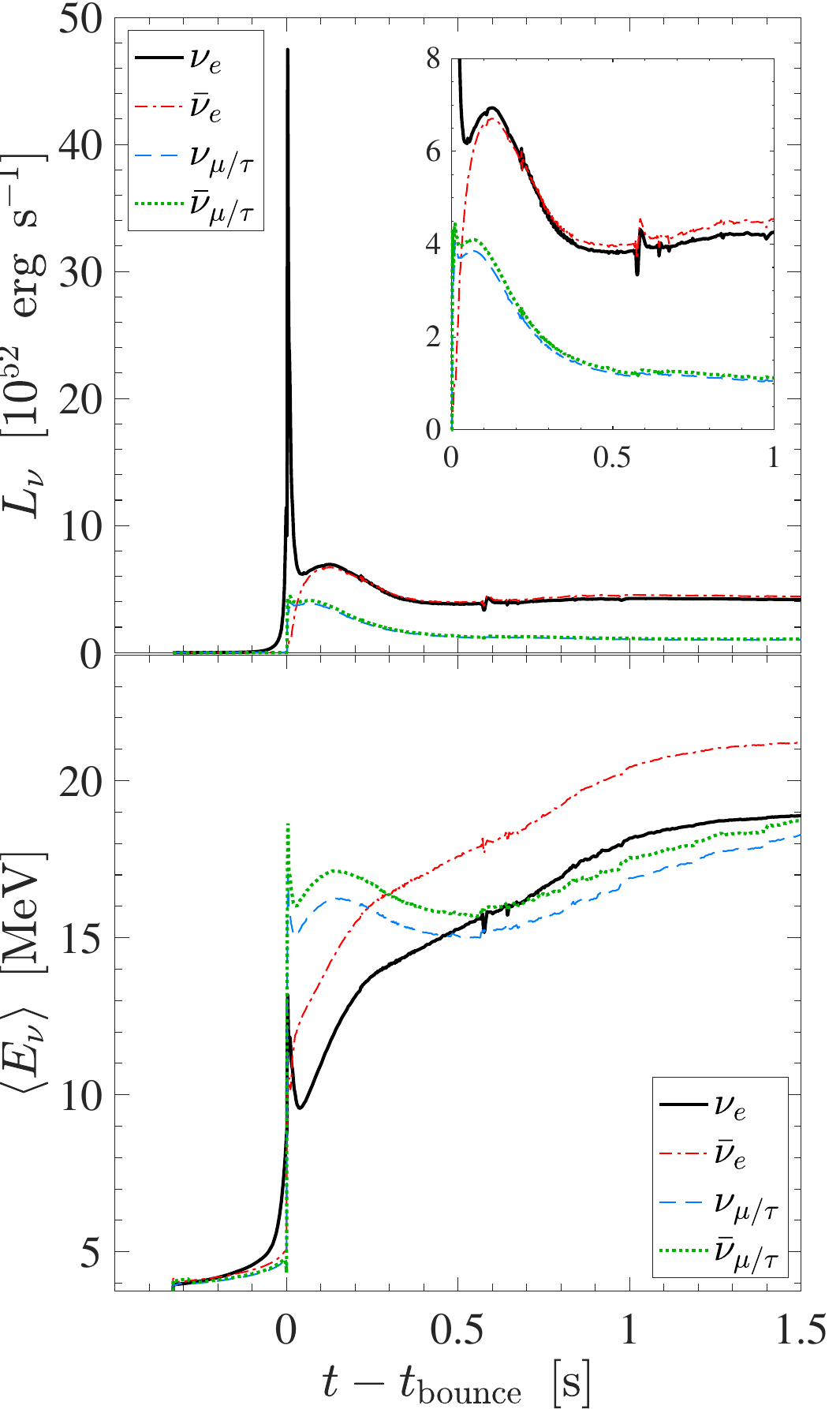}
\label{fig:lumin_a}
}
\subfigure[~DD2-MEV]{
\includegraphics[width=0.4\textwidth]{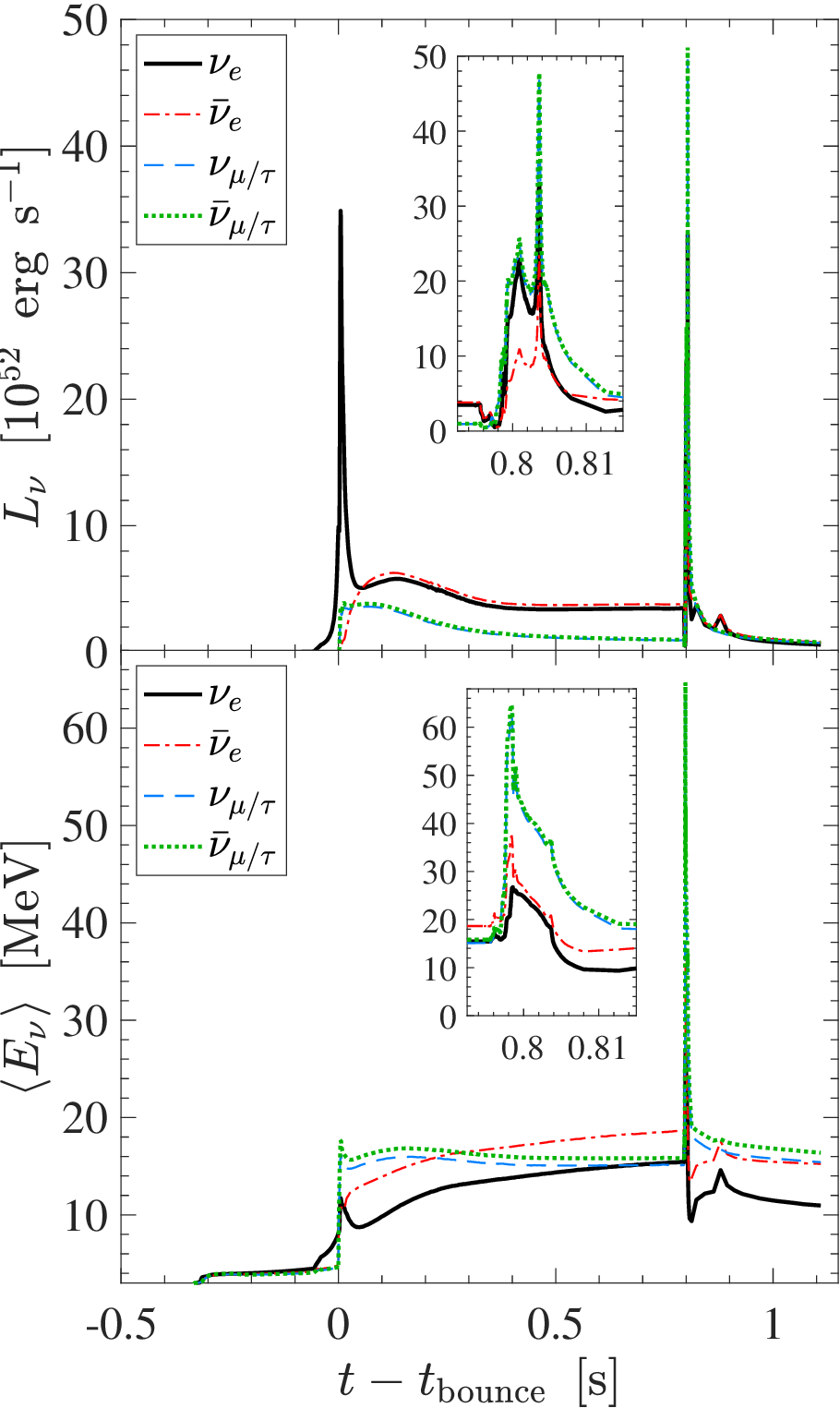}
\label{fig:lumin_b}
}
\caption{Post bounce evolution the neutrino luminosities $L_\nu$ (top panels) and mean energies $\langle E_\nu \rangle$ (bottom panel), for $\nu_e$ (solid black lines), $\bar\nu_e$ (red dash-dotted lines), $\nu_{\mu/\tau}$ (blue dashed lines) and $\bar\nu_{\mu/\tau}$ (green dotted lines), sampled in the co-moving frame of reference at a radius of 500~km.  
\label{fig:lumin}
}
\end{figure}

The neutrino signature from such QCD driven CCSN explosions have long been explored as potential observable signal~\cite{Dasgupta10,Fischer18,Kuroda2022ApJ924_QCD,Khosravi:2024ApJ964}, however, based upon model EOS with generally large phase transition density jumps~\cite{Sagert09,Bastian:2021}, where otherwise no explosions could have been obtained and hence no neutrino signature was identified.
In such cases, the presence of a strong second shock, as a direct consequence of the supersonic PNS collapse due to the first-order phase transition, results in the release of a millisecond neutrino burst during the shock propagation across the neutrinospheres of last inelastic scattering.
Since this is the case here for the run with DD2-MEV, featuring the van der Waals behavior, we confirm the release of a millisecond neutrino burst, qualitatively in agreement with previous investigations (see Fig.~\ref{fig:lumin_b}). 
The neutrino luminosities (see top panel) first drop sharply, as the neutrinosphere radii drop rapidly for all neutrino flavors as the second shock propagates towards the neutrinospheres from below, during which the temperature at the neutrinospheres already rises and hence the average neutrino energies rise accordingly (bottom panel in Fig.~\ref{fig:lumin_a}), reaching peak values on the order of $\langle E_{\nu_e} \rangle=27$~MeV, $\langle E_{\bar\nu_e} \rangle=36$~MeV, and $\langle E_{\nu} \rangle=65$~MeV for $\nu_{\mu/\tau}$ and $\bar\nu_{\mu/\tau}$. Once the shock has passed across the neutrinospheres, also the neutrino luminosities rise sharply of all neutrino flavors. 
However, quantitatively different arise from the previously reported millisecond burst release, due to the substantially longer timescale for formation and propagation of the second shock, also the second neutrino burst is substantially broader (see the inlay in the top panel of Fig.~\ref{fig:lumin_b}). It is divided into a first rise of the  neutrino luminosities, to values of about $L_\nu\simeq 1.0$--$2.5\times 10^{53}$~erg, for about 4--5~ms, due to the slow contraction of the PNS, before the sudden rise associated with the shock passage to even higher values of about $L_{\nu_e}\simeq 3.5\times 10^{53}$~erg, $L_{\bar\nu_e}\simeq 4\times 10^{53}$~erg and $L_{\nu}\simeq 4.8\times 10^{53}$~erg for $\nu_{\mu/\tau}$ and $\bar\nu_{\mu/\tau}$. 
It is interesting to note that during the initial rise of the neutrino luminosities, we find $L_{\bar\nu_e} < L_{\nu_e}$, only during the shock passage across the neutrinospheres, the opposite behavior occurs. 
The later expansion of the second shock across the bounce shock initiates the supernova explosion, after which the shock expands continuously towards increasing radii, and the remnant PNS contracts accordingly due to the deleptonization (see Fig.~\ref{fig:SN-evol_b}). As a consequence, the neutrino luminosities and mean energies start to decrease continuously, on a timescale on the order of 10~s.  

The situation of neutrino emission is qualitatively different for the CCSN with DD2-MEV~(Gibbs). In the absence of a second shock wave, the neutrino signal (see Fig.~\ref{fig:lumin_a}) does not show any millisecond burst-like feature. Instead, as a direct consequence of the structural PNS readjustment during the phase transition---the neutrinospheres closely follow the PNS radius shown in Fig.~\ref{fig:SN-evol_a} (red solid line)---the neutrino luminosities first drop slightly, due to the contracting PNS, and later rise again, due to the expansion of the PNS, back to the magnitude before the phase transition (see the top panel in Fig.~\ref{fig:lumin_b}). The mean energies (bottom panel) follow the same trend. The impact is greatest for $\nu_e$ and $\bar\nu_e$, as their neutrinospheres are located at the lowest densities where the impact of PNS contraction and expansion has the largest impact. After the phase transition, the luminosities and mean energies for DD2-EV~(Gibbs) follow the canonical post-bounce behavior during the mass accretion phase (c.f. Ref.~\cite{Fischer09}), towards black hole formation as is the case of all failed core-collapse supernova explosions.

\section{Gravitational wave mode analysis}
\label{sec:modes}
Besides neutrinos, another potentially observable signal from core-collapse events is provided by oscillations excited in the hydrodynamic and spacetime degrees of freedom of the system. These oscillations encode information about the internal structure and dynamical state of the collapsing object and can manifest, for example, through gravitational-wave emission. In this section, we perform a mode analysis of the CCSN simulations discussed in Sec.~\ref{sec:sim}, following the general relativistic astroseismology approach of Ref.~\cite{Torres-Forne2019MNRAS482}, who employed the numerical {\tt GREAT} code to identify and characterize the dominant oscillation modes excited during the evolution. 
By decomposing the time-dependent fluid and metric variables into their characteristic eigenmodes in spherical symmetry, the analysis links numerical simulation data to the underlying physical processes, such as pressure-driven and buoyancy-driven oscillations, i.e. $p$ and $g$ modes, respectively. Including the fundamental $f$ mode in addition, these can help identify possible observational signatures.

Using the relativistic Brunt-V\"ais\"al\"a frequency and the relativistic Lamb frequency, denoted by $N^2$ and $L^2$, respectively (see Eqs.~(33) and (34) in Ref.~\cite{Torres-Forne2018MNRAS474}), provides a practical way to locate regions where convection is expected to occur. 
The Brunt-V\"ais\"al\"a frequency is linked to the Ledoux stability criterion and depends on gradients in both lepton number and entropy (c.f. Ref.~\cite{mirizzi16} and references therein). 

One can relate $g$ modes to $N^2$, while the $f$ and $p$ modes (which have the same characteristics) are related to $L^2$. Distinct oscillation modes may develop depending on the time evolution of various physical conditions. Gravity modes arise in regions where buoyancy acts as the restoring force, which corresponds to the condition $N^2 > 0$. 
On the other hand, propagation of sound waves excite the $p$ modes in which buoyancy does not play a role. The focus of this study is the interior structure and dynamics of the PNS. Hence, of particular interest are the fundamental and gravity modes, whereas the pressure modes are not considered in the present analysis.
Therefore, for the explicit mode analysis we select as an outer boundary a fixed restmass density of $\rho=10^{11}$~g~cm$^{-3}$, such that most of the $p$ modes are omitted naturally in the analysis. The inner boundary is chosen to be the very central fluid element.
Then, under the assumption of spherical symmetry, the oscillation eigenfunctions are computed by performing a linear, adiabatic perturbative treatment of the coupled hydrodynamic and spacetime field equations around a stationary equilibrium configuration.
The perturbations are taken in the Eulerian framework and decomposed into spherical harmonics for the angular dependence, together with a harmonic time dependence characterized by a single frequency \citep[c.f. Ref.~][]{Torres-Forne2019MNRAS482}.
Within this formalism, the Lagrangian displacement vector of a fluid element can be written as
\begin{eqnarray}
\pmb{\xi} 
&=& 
\bigg [ \eta_r (r) Y_{lm} (\theta,\varphi) \pmb{\hat r} \nonumber \,
+
\,\eta_{\perp}(r)
\left(
\frac{\partial_\theta Y_{lm}(\theta,\varphi)}{r^2} \pmb{\hat \theta}
+
\frac{\partial_\varphi Y_{lm}(\theta,\varphi)}{r^2 \sin^2\theta} \pmb{\hat \varphi}
\right)
\bigg] e^{i \sigma t}~,
\end{eqnarray}
where $(r,\theta,\varphi)$ denote the spherical polar coordinates and $(\pmb{\hat r}, \pmb{\hat \theta}, \pmb{\hat \varphi})$ are the associated orthonormal basis vectors. The functions $Y_{lm}$ represent the usual spherical harmonics.
The parameter $\sigma$ corresponds to the mode angular frequency, while the radial functions $\eta_r(r)$ and $\eta_\perp(r)$ describe the radial and tangential components of the fluid displacement, respectively.
We note that our analysis goes beyond the Cowling approximation, as we consistently include perturbations of both the fluid variables and the spacetime metric. This full treatment of the coupled matter–gravity system is required to accurately capture the global oscillation modes, especially for low-order modes. Throughout this analysis, attention is restricted to the lowest-order gravity mode, corresponding to a single radial node ($n=1$). Higher-order $g$ modes are excluded from consideration.
We include all solutions for modes with an angular degree of $l=2$ within the frequency range up to $10~\mathrm{kHz}$, in numerical steps of 1~Hz. 
In the following nomenclature, we include the fundamental mode, denoted as $f$ mode\footnote{In the literature sometimes denoted as $^2f$ \cite{Torres-Forne2019MNRAS482}, the lower index denotes $n$.}, and the lowest $g$ mode is denoted as $^2g_1$.

\begin{figure}
\centering
\subfigure[~DD2-MEV~(Gibbs)]{
\includegraphics[width=0.66\textwidth]{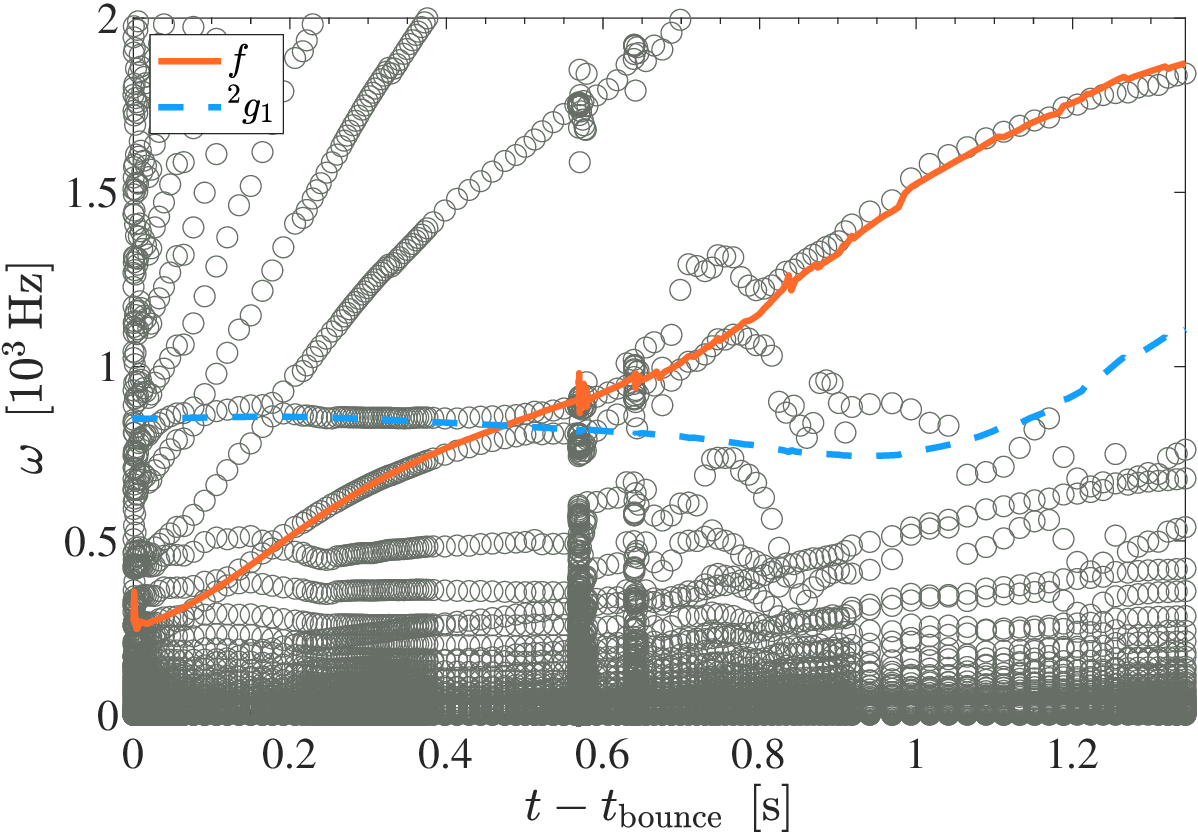}
\label{fig:freqs_Gibbs}
}
\hfill
\subfigure[~DD2-MEV]{
\includegraphics[width=0.66\textwidth]{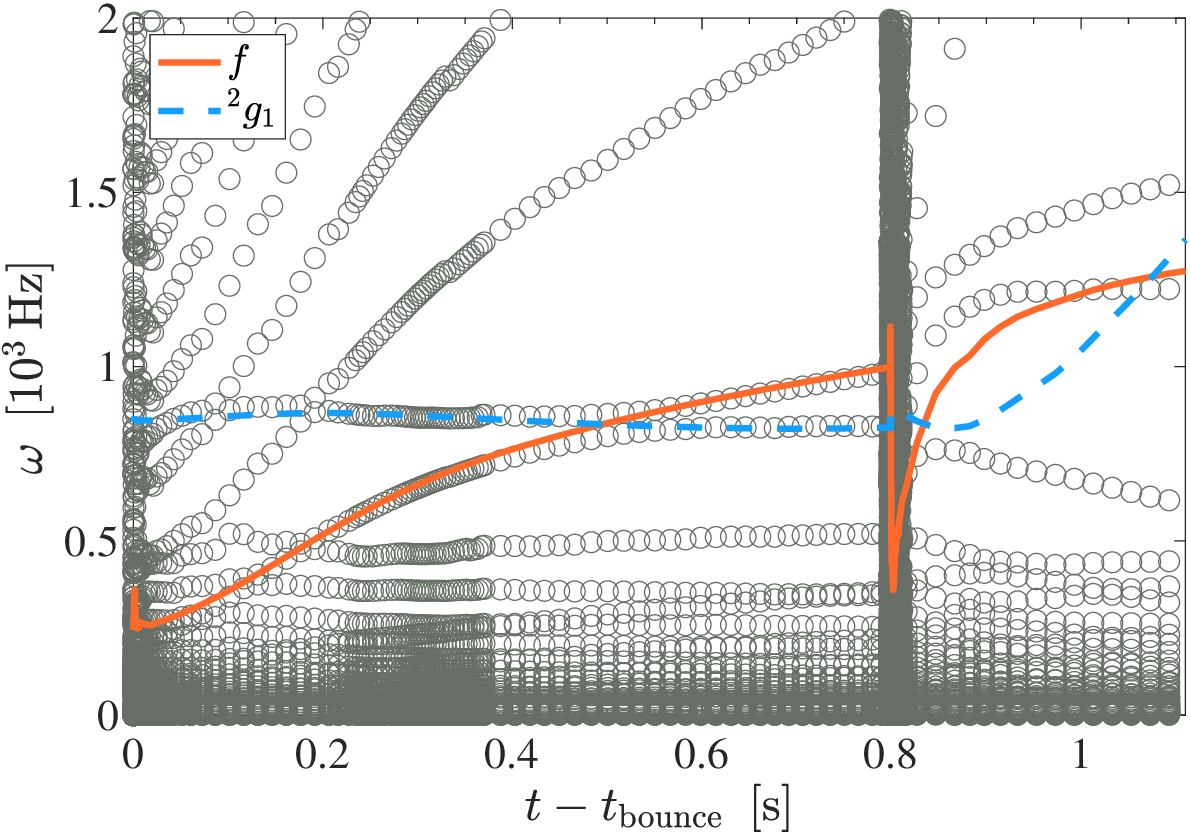}
\label{fig:freq_MEV}
}
\hfill
\subfigure[~DD2F-RDF-1.2]{
\includegraphics[width=0.66\textwidth]{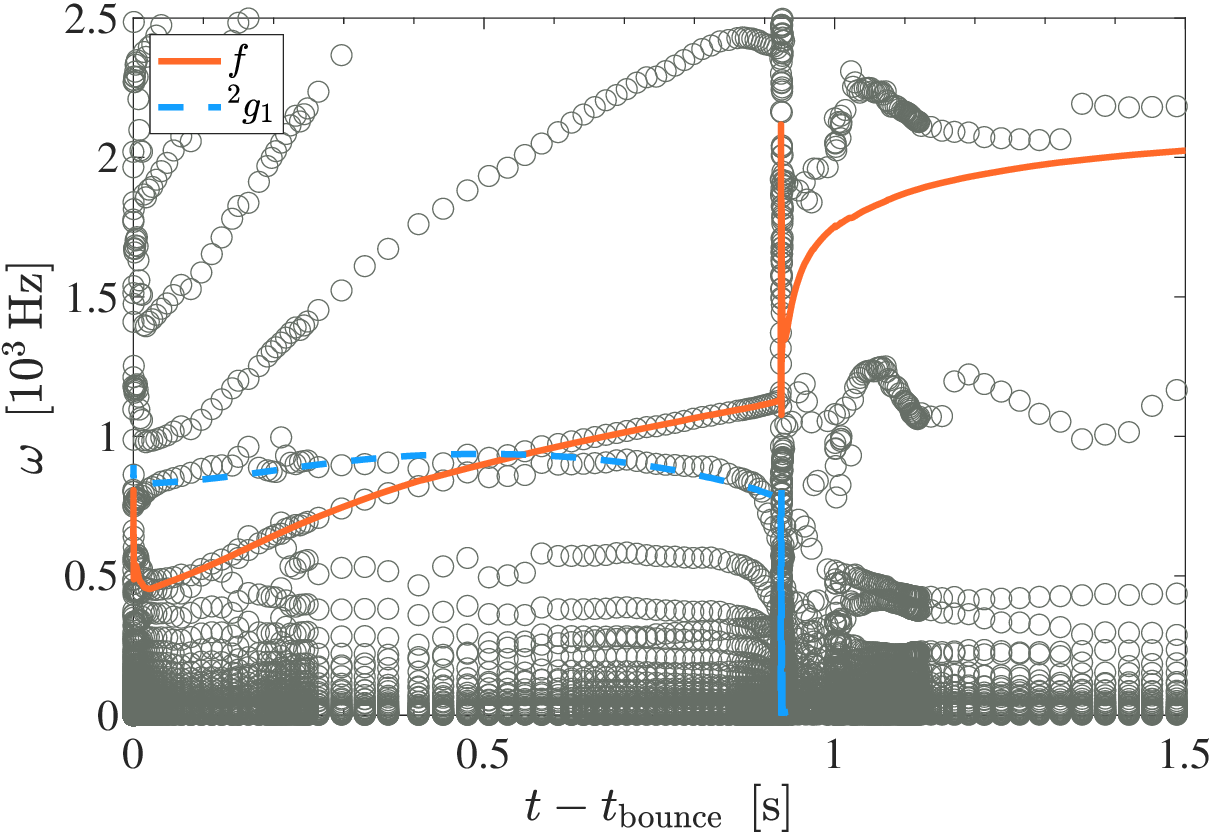}
\label{fig:freq_RDF}
}
\caption{Eigenmode frequency post-bounce evolution obtained from the {\tt GREAT} analysis are shown as gray open circles for the CCSN runs with the DD2-MEV~(Gibbs) EOS, with a phase-transition construction, DD2-MEV, featuring van der Waals phase transition behaviour, and DD2F-RDF-1.2 from Ref.~\cite{Khosravi:2024ApJ964}. 
Fits for $f$ and $^2g_1$ modes are indicated by solid orange and blue dashed lines.
\label{fig:freqs}
}
\end{figure}
\begin{figure}
\centering
\subfigure[~DD2-MEV~(Gibbs)]{
\includegraphics[width=0.31\textwidth]{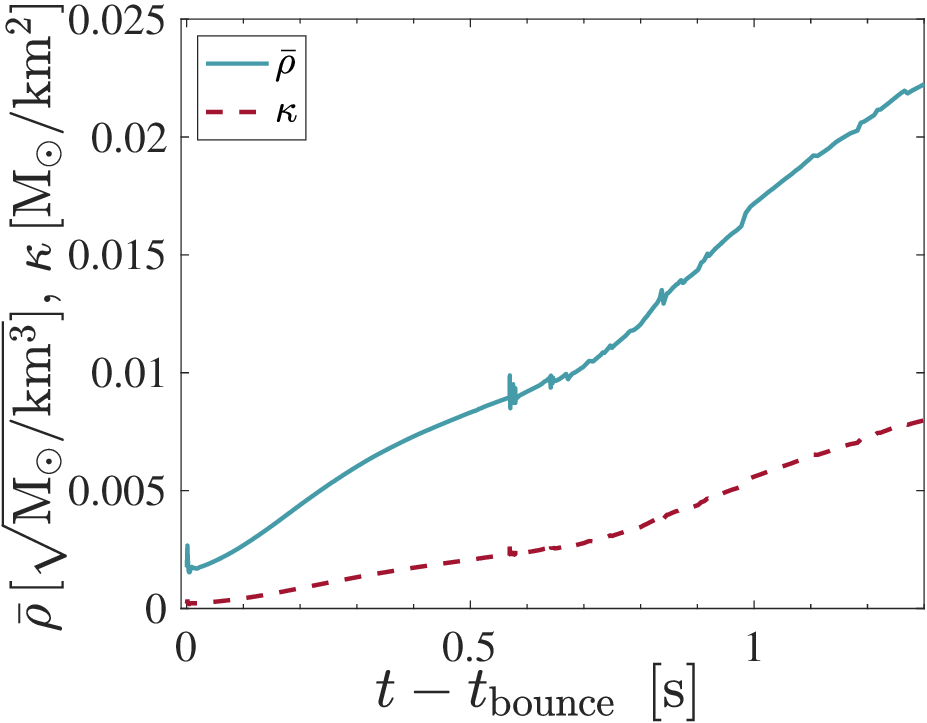}
\label{fig:MR_Gibbs}
}
\subfigure[~DD2-MEV]{
\includegraphics[width=0.31\textwidth]{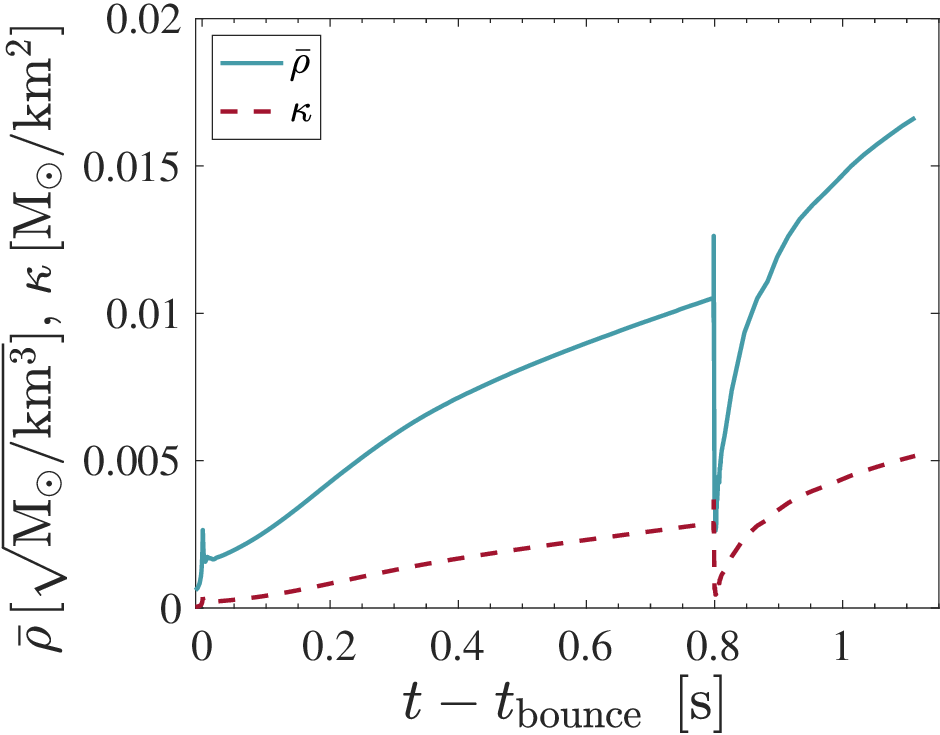}
\label{fig:MR_MEV}
}
\subfigure[~DD2F-RDF-1.2]{
\includegraphics[width=0.31\textwidth]{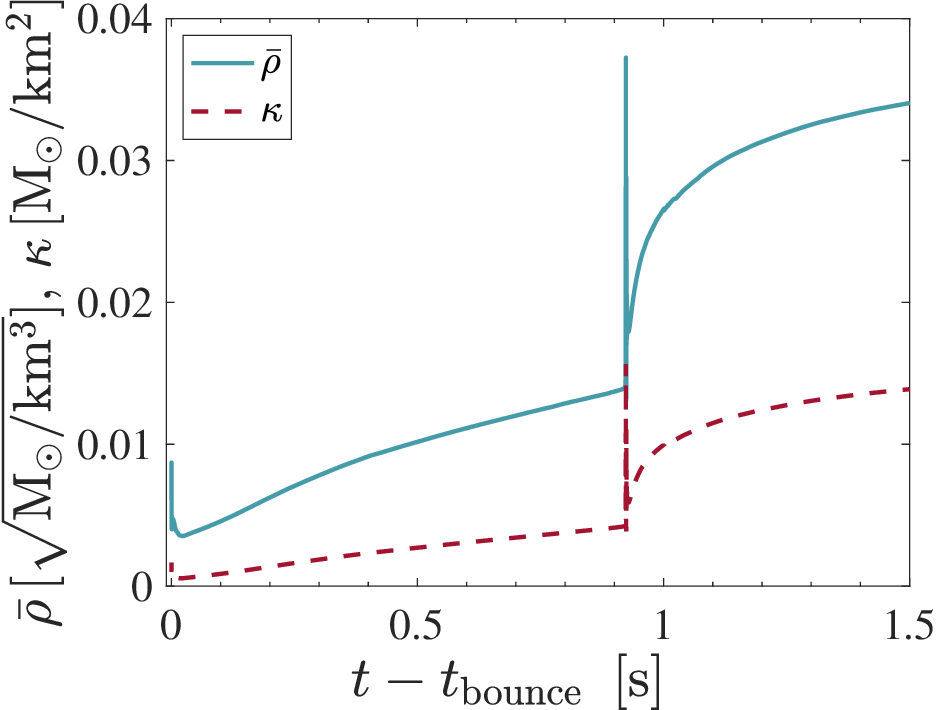}
\label{fig:MR_RDF}
}
\caption{ Post-bounce evolution of PNS compactness, $\kappa$, and mean density, $\bar\rho$, for the CCSN runs with the DD2-MEV~(Gibbs) EOS, with a phase-transition construction, with DD2-MEV, featuring van der Waals phase transition behaviour, and for the DD2F-RDF-1.2 run of Ref.~\cite{Khosravi:2024ApJ964}.
\label{fig:MR}
}
\end{figure}

The mode frequencies obtained from the {\tt GREAT} analysis are shown in Fig.~\ref{fig:freqs}, for the two CCSN simulations with DD2-MEV~(Gibbs) and DD2-MEV EOS, and the mode analysis of DD2F-RDF1.2 EOS from simulations of Ref.~\cite{Khosravi:2024ApJ964} for comparison. 
Due to imposed boundary conditions, most modes associated with pressure oscillations ($p$) are outside this frequency range, i.e. pushed toward higher frequencies. According to the mode classification of Ref.~\cite{Cowling1941MNRAS.101..367C}, the dominant mode, with zero radial nodes, is identified as the fundamental ($f$) mode. It emerges at a frequency of approximately $250~\mathrm{Hz}$, and continues to increase during the CCSN post-bounce evolution. 

The oscillation frequencies of the $f$ and $g$ modes are commonly parameterized in terms of the PNS mass and radius, $M_{\rm PNS}$ and $R_{\rm PNS}$~\cite{Torres-Forne12019PhRvL123}. In particular, the $f$ mode is found to correlate with the mean density, $\bar\rho\equiv\sqrt{M_{\rm PNS}/R_{\rm PNS}^{3}}$, while the $g$ modes are primarily governed by surface gravity, $\kappa\equiv M_{\rm PNS}/R_{\rm PNS}^{2}$. The fitting procedure employed here follows closely Refs.~\cite{Torres-Forne2018MNRAS474,Torres-Forne2019MNRAS482}, where polynomial expansions up to third order are used to describe the mode-frequency relations,
\begin{equation}
\omega_{\rm fit}(x) = a_0 + a_1\,x + a_2\,x^2 + a_3\,x^3~,
\label{eq:fit_CCSN}
\end{equation}
for which the coefficients $a_0$, $a_1$, $a_2$ and $a_3$ are listed in Table~\ref{tab:fit_CCSN_freqs} for the different CCSN runs.
The post-bounce evolution of $\bar\rho$ and $\kappa$ are shown in Fig.~\ref{fig:MR}.

Unlike in canonical CCSN simulations, i.e., without high-density phase transition, when the fundamental mode rises continuously---this reflects the continuously increasing PNS mean density in the presence of mass accretion, either until the CCSN explosion proceeds or until a black hole forms in the case of failed CCSN---the fundamental mode behaves qualitatively different across the phase transition. 
For the DD2-MEV~(Gibbs) simulation, we identify the sudden change of the slope of the $f$ model (see the solid orange line in Fig.~\ref{fig:freqs_Gibbs}, which represents the $f$-mode fit~\eqref{eq:fit_CCSN}, with the parameters given in Tab.~\ref{tab:fit_CCSN_freqs}), as a direct consequence of the structural change of the PNS due to the incomplete phase transition, as was discussed above in Sec.~\ref{sec:sim}. After the PNS has gone through the complete phase transition, $\simeq$100~ms, the PNS features a higher central and mean density (see Figs.~\ref{fig:SN-central_a} and \ref{fig:MR_Gibbs}), and hence the $f$ mode rises faster than before in the pure hadronic phase. 
The $^2g_1$ mode too follows the standard post-bounce behavior before the phase transition~\cite{Torres-Forne12019PhRvL123} (see the dashed blue line in Fig.~\ref{fig:freqs_Gibbs}, which represents the $^2g_1$-mode fit~\eqref{eq:fit_CCSN}, with the parameters given in Tab.~\ref{tab:fit_CCSN_freqs}). However, after the phase transition, the $^2g_1$ modes starts to decrease. This behaviour is observed in only a few cases featuring hadronic EOS, at late times during the post-bounce evolution~\cite{Torres-Forne2019MNRAS482}, however, not as rapidly as here. The DD2-MEV~(Gibbs) $^2g_1$-mode distinguishes qualitatively from those of canonical hadronic EOS towards black hole formation.
Furthermore, the $^2g_1$ mode follows the $\kappa$-dependence only until the phase transition. This indicates the evolution of the $^2g_1$ mode is governed by a different phenomenology in hadronic and quark matter. 
We note that the moment of black hole formation is excluded in the analysis, since the limitation of a static background of {\tt GREAT} becomes invalid with the timescale shortening to $\sim$$10^{-9}$~s, indicated by the lapse function decreasing to $\alpha\simeq 0.1$ (see Fig.~\ref{fig:SN-evol_a}). 

In contrast, the explosion model based on DD2-MEV, with van der Waals phase transition, illustrated in Fig.~\ref{fig:freq_MEV}, both $f$ and $^2g_1$ modes show quantitatively the same mode behaviour only until the moment of phase transition. At the moment of PNS contraction and formation of the second shock, about 0.8~s post bounce, the values of the $f$-mode and $^2g_1$-mode frequencies fluctuate largely, reflecting the large scale oscillations of PNS central and mean density as well as PNS compactness (see Fig.~\ref{fig:MR_MEV}). However, similar to what was reported for the $^2g_1$-mode behaviour for DD2-MEV~(Gibbs), also here  for DD2-MEV the behaviour of both $f$- and $^2g_1$-modes cannot be accurately captured by the dependencies on PNS mean density and compactness (see the red solid and blue dashed lines for the $f$- and $^2g_1$-mode fits, respectively). This indicates that these modes are governed by other physical phenomena.

We confirm these finds as general, by performing a {\tt GREAT} mode analysis for the CCSN simulation with DD2F-RDF-1.2 EOS from the models of Ref.~\cite{Khosravi:2024ApJ964}, launched from the same {\tt s40a28} progenitor model. The results are illustrated in Fig.~\ref{fig:freq_RDF}, together with the corresponding fits for $f$-- and $^2g_1$--mode frequencies (the post-bounce evolution of $\bar\rho$ and $\kappa$ are given in Fig.~\ref{fig:MR_RDF}). 
The values of the parameters are given in Table~\ref{tab:fit_CCSN_freqs}, which differ slightly from those of the DD2-MEV runs, as the DD2F hadronic EOS is slightly softer than DD2, i.e. before the appearance of quark matter. Nevertheless, also for DD2F-RDF-1.2 it becomes evident that the fitting relations~\eqref{eq:fit_CCSN} for $f$ and $^2g_1$ modes break down and fail to reproduce correctly the behaviour of these modes after the first order QCD phase transition (see Fig.~\ref{fig:freq_RDF}).

\begin{table}[t!]
\centering
\caption{Parameters for the frequency fits expression~\eqref{eq:fit_CCSN}.}
\begin{tabular}{c c c c c c c}
\hline
EOS & mode & $x$ & $a_0$ & $a_1$ & $a_2$ & $a_3$ \\
& & & & $[10^4]$& $[10^6]$& $[10^{9}]$\\
\hline
\hline
DD2-MEV (Gibbs) & $f$ & $\bar\rho$ & $100$ & $9.75$ & $-0.85$ & $0$\\
DD2-MEV (Gibbs) & $^2g_1$ & $\kappa$ & $844$ & $-3.52$ & $27.71$& $3.34$ \\
\hline
DD2-MEV & $f$ & $\bar\rho$ & $75.53$ & $11.45$ & $-2.55$ & $0$\\
DD2-MEV & $^2g_1$ & $\kappa$ & $826.35$ & $10.75$ & $-84.30$& $16.27$ \\
\hline
DD2F-RDF-1.2 & $f$ & $\bar\rho$ & 192.03 & 7.64 & -0.66 & 0  \\
DD2F-RDF-1.2 & $^2g_1$ & $\kappa$ & 820.78 & -0.74  & 53.23  & -12.80  \\
\hline
\end{tabular}
\label{tab:fit_CCSN_freqs}
\end{table}
%

\section{Summary and conclusions}
\label{sec:summary}
The present paper overcomes one major caveat of previous core-collapse supernova studies featuring a first-order QCD transition, which were exclusively based on the two-phase approach. To this end, we adopt the excluded volume method of Ref.~\cite{Typel2018Univ4}. It extends the DD2 RMF model EOS featuring the change of degrees of freedom via a modified excluded volume functional, which mimics van der Waals behaviour with over-compressed and under-compressed regions.
A similar approach has been adopted recently to the neutron star phenomenology in~\cite{2025arXiv251208672A} based on the polytropic EOS. 

This novel class of DD2-MEV EOS is then implemented in simulations of core-collapse supernovae in spherical symmetry, featuring general relativistic neutrino radiation hydrodynamics and six-species Boltzmann neutrino transport. The latter is essential in enabling us to make reliable predictions of the associated neutrino signal, in particular the possible  release of a burst-like signature from the QCD phase transition, as was previously predicted.
In contrast with previously reported successful supernova explosions, driven by a first order hadron-quark matter phase transition, featuring the two-phase approach and phase transition construction within the relativistic density functional formulation, explosions are obtained here only for the DD2-MEV model with van der Waals behavior. For the DD2-MEV~(Gibbs) with phase transition construction, the density jump between hadron and quark matter phases was substantially smaller than those of the RDF hybrid models~\cite{Kaltenborn17,Bastian:2021}, in particular the strong temperature dependence of the onset density of Refs.~\cite{Fischer18,Bastian:2021,Fischer:2021EPJA57} could not be reproduced within the class of DD2-MEV EOS. 
This is particularly critical as it has been demonstrated recently that the density functional of the RDF models contains a scalar interaction contribution that mimics confinement, however, its density dependence cannot be related to QCD~\cite{ShuklaPok2025JSPC300058S}. It has also been pointed out recently that such mean field models of quark matter fail to fulfill the asymptotic behavior of QCD apriori, well investigated within perturbative QCD~\cite{Kurkela:2014vha,Kurkela:2020NatPh}, i.e. the speed of sound diverges, for which non-local interactions with momentum dependent interactions have been proposed as potential solution to this problem (c.f. Ref.~\cite{ShuklaPok2025arXiv250706741S}, and references therein).

The release of a neutrino burst is hence absent in the simulation based on DD2-MEV~(Gibbs), only short-time fluctuations on the neutrino fluxes and mean energies could be observed, as a direct consequence of the structural transition of the PNS in the hadron-quark matter mixed phase. Once pure quark matter has been reached, the later post-bounce evolution proceeds without shock revival towards black hole formation. 
The neutrino burst released in the DD2-MEV model is found to be substantially broader, on the order of 5--10~ms, compared to previously reported 1--2~ms bursts released, based on MIT bag models~\cite{Sagert09,Fischer11} as well as RDF EOS~\cite{Fischer18,Khosravi:2024ApJ964}~\cite{Sagert09,Fischer11,Fischer18,Fischer:2021EPJA57}. 
This is associated here with the delay in the second shock formation and expansion. The resulting initial ultra-relativistic explosion is found to become substantially dimmer towards several 100~ms after the explosion onset, related to massive fall back. 
Reliable asymptotic values of the explosion energy estimate could not be obtained at this stage. Therefore, the simulations will have to be carried out for longer times. Nevertheless, these results highlight the large differences compared to previously reported, nearly ballistically-driven CCSN explosions driven by a first-order QCD phase transition, and, moreover, the sensitivity of the results to details of the underlying quark-hadron hybrid EOS and the phase transition itself. 

Furthermore, we performed a detailed gravitational wave mode analysis, which reveals distinct difference of the behaviours of fundamental and lowest gravity modes, between failed CCSN and explosions, both of which feature first-order phase transitions, with phase transition construction and continuous van der Waals behaviour, respectively. This might enable us to distinguish whether quark matter appears in failed CCSN from purely hadronic events, if confirmed within multi-dimensional simulations, which we leave for future explorations. 
Morerover, we find that the commonly explored dependencies of $f$ and $g$ modes, in terms of PNS mean density and compactness, give similar fitting relations in terms of these valiabels for the $f$ modes, however, largely different fit parameters for the $^2g_1$ modes, for the three CCSN models explored here, i.e. DD2-MEV~(Gibbs)---failed CCSN explosion and black hole formation---, DD2-MEV---first-order phase transition driven explosion---and DD2F-RDF-1.2~\cite{Khosravi:2024ApJ964}. We also note that the fitting procedure in terms of mean density and surface compactness fails to provide a qualitative explanation for the behaviour of these modes after the appearance of quark-matter degrees of freedom, indicating a different physical origin for the excitation of these modes. 
We note further that the behaviour of the eigenmodes during the extended phase transition evolution for DD2-MEV, which is significantly longer than for previously explored CCSN models, will have to be considered in future more detailed investigations, due to the imaginary speed of sound and the associated implication for the appearance of hydrodynamical instabilities~\cite{ChomazColonnaRandrup:2004PhR389}, including consequences for the physical interpretation and determination of Brunt-V\"ais\"al\"a and Lamb frequencies. In the present analysis, we avoid spurious solutions of the eigenvalue problem by nulling the speed of sound in the region of DD2-MEV instability. 

In order to confirm these novel results based on the continuous EOS behaviour of the hadron-quark matter phase transition, featuring van der Waals behavior, it will be essential to explore additional parameterization of the class of DD2-EMV EOS. This concerns, in particular, configurations with a wider density jump for the critical regions as well as deeper and more shallow criticlalities, all in contrast with the cases with phase transition construction, as was discussed in the present paper for one particular parameterization. 
The results reported here remain to be verified in multi-dimensional general relativistic CCSN simulations, which allow to directly compute the gravitational wave signals, which we leave for future explorations.

\data{The multi-purpose DD2-MEV equation of state tabulations generated in this study are available at the authors based upon reasonable request. Supernova simulation datasets generated during this work can be obtained from the corresponding author upon reasonable request. All scripts and codes used to produce the tabulations and simulations are included in the repository.}

\section*{Acknowledgments}
The authors thank Hans-Thomas Janka, Micaela Oertel and Fiorella Burgio for fruitful discussions. 
We thank the organizers of SN2025GW---First IGWN Symposium on Core Collapse Supernova Gravitational Wave Theory and Detection, which was held in Warsaw in July 2025.
The supernova simulations were performed at the Wroclaw Center for Scientific Computing and Networking (WCSS).

\section*{Conflict of interest}
The author declares no conflicts of interest.

\section*{Data access statement}
This work generates multi-purpose equation of state tabulations, which will be made publicly available upon publication. Supernova simulation data were also generated as part of this study and will be shared upon reasonable request. Detailed instructions for accessing these datasets, including any associated software or scripts, will be provided in the supplementary materials.

\section*{Ethics statement}
No ethical issues arise, as no test subjects are involved. This paper adheres to academic integrity.

\section*{Funding statement}
This work was supported by the Polish National Science Center (NCN) under Grants Nos. 2023/49/B/ST9/03941 (A.K. and T.F.) and 2023/49/N/ST9/03995 (N.K.L.).

\appendix

\section*{Appendix: Phase transition construction}
\label{app:Gibbs}
In order to construct the phase transition, we have to interpolate linearly the (extensive) thermodynamic potential, that is appropriate for the used independent variables temperature, $T$, charge chemical potential, $\mu_{Q}$\footnote{In the absence of other fermionic hadrons, $\mu_Q = \mu_p - \mu_n$, which is the difference between proton and neutron chemical potentials, $\mu_p$ and $\mu_n$, respectively.}, and baryon density, $n_{\rm B}$. We select the new thermodynamic potential $\tilde{F}(T,n_{\rm B},\mu_Q) = F(T,n_{\rm B},n_Q)-\mu_Q n_Q$, which is obtained by a Legendre transformation of $F$ changing $n_{Q}$ to $\mu_{Q}$ \cite{Typel2014EPJA50}. 

We thus reduce the problem to an effective one-dimensional Maxwell construction at fixed temperature $T$ and charge chemical potential $\mu_Q$, with the baryon number density $n_B$ as a single extensive variable. Then the densities can be easily determined at the two boundaries of the coexistence region 
respecting the general Gibbs conditions, i.e., the equality of the intensive thermodynamic variables at the low- and high-density points of coexistence.
We already know that
\begin{equation}
\label{eq:pdef}
    -p = \left. \frac{\partial \tilde{F}}{\partial V} \right|_{T,N_{\rm B},\mu_{Q}} 
    = -n_{\rm B}^{2} \left. 
    \frac{\partial \left( \tilde{f}/n_{\rm B} \right) }{\partial n_{\rm B}} \right|_{T,\mu_{Q}} 
    \quad \mbox{with} \quad \tilde{f}=\frac{\tilde{F}}{V}~, \quad n_{\rm B} = \frac{N_{\rm B}}{V}~.
\end{equation}
Thus for constant pressure $p$, we can integrate equation (\ref{eq:pdef}) and obtain
\begin{equation}
    \frac{\partial \left( \tilde{f}/n_{\rm B} \right)}{\partial n_{\rm B}} = \frac{p}{n_{\rm B}^{2}}
    \qquad \Rightarrow \qquad 
    \frac{\tilde{f}}{n_{\rm B}} = -\frac{p}{n_{\rm B}} + C
    \qquad \Rightarrow \qquad
    \tilde{f}(n_{\rm B}) = -p + C \, n_{\rm B}
\end{equation}
with an integration constant $C$ for a general baryon density $n_{\rm B}$ with $n_{\rm B}^{I} \leq n_{\rm B} \leq n_{\rm B}^{II}$  between the boundary densities, $n_{\rm B}^{I}$ and $n_{\rm B}^{II}$, of the coexistence region. 
Then, the modified free energy density can be expressed as follows,
\begin{equation}
    \tilde{f}(n_{\rm B}) = (1-\chi) \tilde{f}(n_{\rm B}^{I}) + \chi \tilde{f}(n_{\rm B}^{II})
    \qquad \Rightarrow \qquad
    C = \frac{\tilde{f}(n_{\rm B}^{II})-\tilde{f}(n_{\rm B}^{I})}{n_{\rm B}^{II}-n_{\rm B}^{I}}
\end{equation}
with
\begin{equation}
    \chi = \frac{n_{\rm B}-n_{\rm B}^{I}}{n_{\rm B}^{II} - n_{\rm B}^{I}}~.
\end{equation}

Note that the intensive quantities, $p$, $T$, $\mu_{\rm B}$, and $\mu_{Q}$, are constant in the coexistence region, when $\tilde{f}$ is used with its proper variables. 
The entropy density can then be found in this region by a linear interpolation as follows,
\begin{equation}
    s(n_{\rm B}) = \frac{S}{V} 
    = - \left. \frac{\partial (\tilde{F}/V)}{\partial T}\right|_{V,N_{\rm B},\mu_{Q}}
    = - \left. \frac{\partial \tilde{f}}{\partial T} \right\vert_{n_{\rm B},\mu_{Q}} =
    (1-\chi) s(n_{\rm B}^{I}) + \chi \, s(n_{\rm B}^{II})
    \: .
\end{equation}
Furthermore, since the charge number is an extensive quantity, a linear interpolation is applied for the charge or isospin density, similar to that for the modified free energy and for the entropy density.
Finally, we map the resulting phase boundaries to the $n_{\rm B} - Y_p$ plane for constant temperatures.

\bibliography{references}
\bibliographystyle{iopart-num}

\end{document}